\documentclass[epj,nopacs]{svjour}
\usepackage{graphicx}
\usepackage{color}
\usepackage{latexsym}
\usepackage[colorlinks,linkcolor=blue,citecolor=blue,urlcolor=black]{hyperref}
\usepackage{ulem}
\usepackage{dcolumn}% Align table columns on decimal point
\usepackage{bm}% bold math
\usepackage{epsfig}
\usepackage{amsmath}
\usepackage{amssymb}
\usepackage{subfigure}
\usepackage{float}
\usepackage{ulem}

\def \a{\alpha}
\def \b{\beta}

\def \d{\Delta}

\def \a{\alpha}

\def \th{\theta}

\def \f{\frac}

\newcommand{\be}{\begin{eqnarray}}
\newcommand{\ee}{\end{eqnarray}}

\begin{document}
\title{Investigating the existence of gravitomagnetic monopole in M87*}
\author{M. Ghasemi-Nodehi\inst{1}\inst{2}\thanks{\emph{email:} mghasemin@ipm.ir}, Chandrachur Chakraborty\inst{3}\inst{4}\thanks{\emph{email:} chandrachurchakraborty@gmail.com}, Qingjuan Yu\inst{3}\thanks{\emph{email:} yuqj@pku.edu.cn}, \& Youjun Lu\inst{1}\inst{5}\thanks{\emph{email:} luyj@nao.cas.cn}% etc
%\thanks is optional - remove next line if not needed
%
}                     % Do not remove
\institute{National Astronomical Observatories, Chinese Academy of Sciences, Beijing 100101, China
\and School of Astronomy, Institute for Research in Fundamental Sciences, Tehran 19395-5531, Iran
\and Kavli Institute for Astronomy and Astrophysics, Peking University, Beijing 100871, China
\and Department of Physics, Indian Institute of Science, Bengaluru 560012, India
\and School of Astronomy and Space Science, University of Chinese Academy of Sciences, Beijing 100049, China
}
\date{Received: date / Revised version: date}
\abstract{
We examine the possibility for the existence of gravitomagnetic
monopole ($n_*$) in M87* by using the results obtained from its first
Event Horizon Telescope image. By numerically deducing the shadow sizes in Kerr-Taub-NUT (KTN)
spacetime, we show that the shadow size increases with increasing $|n_*|$ for a
fixed Kerr parameter $|a_*|$ in case of the KTN black hole, whereas for a KTN naked singularity it increases with increasing $n_*$ for a fixed $a_* > 0$  if $n_* > -\cot 17^{\circ}$ . In general, the asymmetry of shadow shape increases if the central dark object in M87 is a KTN/Kerr naked singularity instead of
a KTN/Kerr black hole. We find that a
non-zero gravitomagnetic monopole is still compatible with the current EHT
observations, in which case the upper limit of $n_*$ cannot be greater than $1.1$, i.e., $n_* \lesssim 1.1$ for the prograde rotation ($a_* > 0$), and the lower limit
of $n_*$ cannot be less than $-1.1$, i.e., $ n_* \gtrsim -1.1$ for the retrograde rotation ($a_* < 0$).
Moreover, if the circularity of the shadow can be measured on a precision of $\lesssim 1\%$,
the Kerr and KTN naked singularities can be falsified for M87*.
\PACS{
      {PACS-key}{discribing text of that key}   \and
      {PACS-key}{discribing text of that key}
     } % end of PACS codes
} %end of abstract
\authorrunning{M. Ghasemi-Nodehi, et al.}
\maketitle

\section{Introduction}\label{intro}
The Kerr spacetime is a stationary and axisymmetric vacuum
solution of the Einstein field equation and it is described by only two
parameters: mass and spin of the collapsed object. Although other axisymmetric
vacuum solutions of the Einstein equation do exist, the most prominent solution
among all of them is the Kerr geometry from the astrophysical point of view.
The Event Horizon Telescope (EHT) has also recently mapped the central compact
radio source of the elliptical galaxy M87 to a Kerr black hole with
unprecedented angular resolution \cite{EHT1}. Adopting the working hypothesis
that M87 contains a Kerr black hole (see Sec. 7.4 of \cite{EHT5}), i.e.,
spin parameter $-1 < a_* < 1$ (see Sec. 1 of \cite{EHT5}), the EHT
collaboration has tried to show that the observed image is overall consistent
with the expectations for the shadow of a Kerr black hole \cite{EHT1}.
However, the alternatives to the Kerr BH have not been ruled out, and
it has been suggested \cite{EHT5} to consider whether the data is also
consistent or not with alternative models for the central object of M87. For
example, it is suggested that the BHs with NUT (Newman-Unti-Tamburino) charges
\cite{nut} could also be possible \cite{EHT5} instead of a Kerr BH. One primary
purpose of the present paper is to show how the NUT charge affects the shadow
size and shape and whether the existence of the NUT charge, aka gravitomagnetic
monopole, can be ruled out or not in the central compact radio source of the
elliptical galaxy M87. To show this, we use the
observational parameter values of the first image of M87*, as released by
the EHT collaboration. The second and most important purpose of this paper is
to constrain the values of the Kerr parameter and the NUT parameter of M87*, if it contains the non-zero gravitomagnetic monopole.

It was argued that the Kerr superspinar (or Kerr naked singularity $|a_*| > 1$
\cite{ckj,ckp}) model for M87* is ruled out by the EHT2017 observations
\cite{EHT5} since the shadows of Kerr naked singularities are substantially
smaller and very asymmetric compared to those of Kerr BHs (see Sec. 8 of
\cite{EHT1}), although other alternatives to the Kerr BH are not ruled out.
However, it has recently been shown in \cite{sc} that the inferred
circularity and size of the shadow of M87* do not exclude the possibility that
this object might be a superspinar. Therefore, one cannot conclude whether M87*
is a Kerr BH ($-1 < a_* < 1$) or a Kerr superspinar ($|a_*| > 1$). Now,
considering the `no body in nature is exactly nonrotating' \cite{jh}, if we
want to test the existence of NUT charge in M87*, we should use the more
general Kerr-Taub-NUT (KTN) spacetime instead of the Kerr spacetime. Referring
to the recent work by \cite{sc}, here we also do not exclude the
possibility that M87* could be a KTN naked singularity (NS) \cite{wei}. Thus,
our `test' includes both the KTN BHs and NSs.

The KTN spacetime is a stationary and axisymmetric vacuum \footnote{Here the vacuum is defined by the vanishing of the symmetric part of the Einstein tensor. Note that the antisymmetric
part of the Einstein tensor of the KTN spacetime does not vanish along the axisymmetric pole (e.g., see the Einstein-Cartan theory beyond the classical general relativity; \cite{hehl}.} solution of the Einstein equation. 
As mentioned in \cite{EHT5}, the KTN BH is within general
relativity with an additional field, i.e., the Einstein-Hilbert action requires
no modification \cite{rs2} to accommodate the NUT charge. Thus, the KTN
solution is related to neither merely post Newtonian nor some modified theory
\cite{rs,rs2}. We note that Bonnor \cite{bon} physically interpreted this NUT
charge as `a linear source of pure angular momentum' \cite{dow,rs}, i.e., `a
massless rotating rod', which is a fundamental aspect of physics \cite{rs2}. If
the NUT charge vanishes, the KTN spacetime reduces to the Kerr spacetime.
Similarly, if the Kerr parameter vanishes, the KTN spacetime reduces to the
Taub-NUT spacetime which includes only two parameters : mass and NUT charge.
% If the NUT charge vanishes, the KTN spacetime reduces to the Kerr spacetime. Similarly, if the Kerr parameter vanishes, the KTN spacetime reduces to the
% Taub-NUT spacetime which includes only two parameters : mass and NUT charge. Bonnor \cite{bon} physically interpreted this NUT
% charge as `a linear source of pure angular momentum' \cite{dow,rs}, i.e., `a
% massless rotating rod', which is a fundamental aspect of physics \cite{rs2}. 

One intriguing feature that emerges here is, the Taub-NUT metric is not
asymptotically flat \cite{mis} in the sense that coordinates cannot be introduced
for which $g_{\mu\nu}-\eta_{\mu\nu}=\mathcal{O}\left( 1/r \right)$. The Taub-NUT spacetime is not asymptotically flat as it contains a string of torsion that extends to the infinity, which is
beyond the classical general relativity and sourced by the NUT charge \cite{hehl}. On the
other hand, the Taub-NUT space is asymptotically flat \cite{mis} in the sense
that the Riemann tensor vanishes ($R_{\mu\nu\a\b} = \mathcal{O}\left( 1/r^3
\right)$) for $r \rightarrow \infty$ as the Schwarzschild case. Note
particularly that the curvature components, and therefore all invariants formed
from the Riemann tensor, depend only on $r$, and not on the other coordinates
\cite{mis}. Thus, the Taub-NUT metric is `an
asymptotically zero curvature space which apparently does not admit
asymptotically rectangular coordinates' \cite{mis}. Specifically, while the
time-coordinate slices are intrinsically asymptotically flat \cite{bini,zs},
the fact that $g_{t\phi} \rightarrow -2n\cos\th$  as $r \rightarrow \infty$ (see Eq. 1 below)
implies that the spacetime is not asymptotically flat. The non-vanishing
$g_{t\phi}$ term leads to the anisotropy at $r \rightarrow \infty$ due to the
presence of $n$, which is also the reason that the Taub-NUT metric may describe a homogeneous but not isotropic cosmological model \cite{mis}. In reality, we consider our Universe as homogeneous and isotropic.
Therefore, if any astrophysical object contains the gravitomagnetic monopole,
in order to nullify its effect at infinity or in order to intact the character
of our Universe as `homogeneous and isotropic', other astrophysical object(s)
of our Universe should also contain gravitomagnetic monopole, so that the total
effects of gravitomagnetic monopoles can vanish.
However, the Taub-NUT metric asymptotically coincides with the leading
approximation for large $r$ of a KTN space \cite{pnas} with electric mass $M$
and magnetic mass $n_*$. 
The asymptotic structures of the Taub-NUT and KTN
spacetimes have recently been analysed in details in Sec. IV C of \cite{bun}
and Sec. IV of \cite{vir} respectively.

Note that the time coordinate in the Taub-NUT metric would be discontinuous at the axisymmetric pole, and the
spacetime contains close timelike curves around the axisymmetric pole, which would raise the causality violation issue \cite{rs2,kag}. Misner \cite{mis} argued that the discontinuity in time could be eliminated by making the time coordinate periodic, but
a periodic time would not describe reality. In this paper, we apply the Novikov self-consistency principle/conjecture (only
self-consistent trips back in time would be permitted \cite{nov,fn}) to avoid the causality violation for the phenomena occurring
in M87*.

Interestingly, it was argued in \cite{pnas} that a nonrotating black hole may be
set in rotation through successive throwing of electric and magnetic monopoles
into it, and after completion of this sequence of processes, a Kerr collapsed
object could be formed. Performing the similar analysis on Taub-NUT
space with only magnetic mass $n_*$, one can set it in rotation and a KTN
collapsed object could be formed (see Conclusions of \cite{pnas}). It was also
shown in \cite{pnas} that these systems, as seen from large distances, are
endowed with an angular momentum proportional to the product of the two kinds
of charges/masses. Remarkably, this angular momentum associated with the
charge-monopole/electric mass-magnetic mass system finally loses all traces of
its exotic origin and is perceived from the outside as `common rotation'.
Therefore, it is perhaps not totally inconceivable to think that, at least part
of the rotation of some of the observed compact objects of our Universe, might
come from their hiding `magnetic poles' which have not yet been observed
\cite{pnas}.

It was once suggested in \cite{lnbl} that the signatures of gravitomagnetic
monopole aka NUT charge might be found in the spectra of supernovae, quasars,
or active galactic nuclei (see also \cite{kag,liu,cc}). However, the
observational evidence of this aspect of fundamental physics was elusive. In a
very recent paper, the first observational indication of the gravitomagnetic
monopole has been reported \cite{cbgm}, based on the X-ray observations of an
astrophysical collapsed object: GRO J1655-40 and it has been shown there that
the accreting collapsed object GRO J1655-40 could be better described with the
more general KTN spacetime \cite{cbgm2}, instead of the Kerr spacetime. Now, as
the compact radio source at the core of the galaxy M87 forms the primary
component of an active galactic nucleus (AGN), one could expect the existence
of the gravitomagnetic monopole in M87* too. This also motivates us to hunting
the existence of gravitomagnetic monopole in M87*.

One should note here that the shadow structures for the various spacetimes are recently investigated in several papers. For example, the shadow structures of a Kerr-like wormholes \cite{1Amir}, charged wormholes \cite{3Amir}, 5D electrically charged Bardeen black holes \cite{4Jusufi}, uncharged \cite{2Ghosh} and charged \cite{6Kumar} rotating regular black holes are thoroughly investigated. The shadow structures of a rotating black holes in 4D Einstein-Gauss-Bonnet gravity is also discussed in \cite{5Kumar}. The parameter estimation of different kinds of black holes have also been illustrated by constraints from the black hole shadow \cite{7Kumar,afrin}. 

The scheme of the paper is as follows. In Sec. \ref{sec:1}, we briefly
describe the KTN spacetime. We outline the basic structure to study the BH
shadow  in Sec. \ref{sec:2} and apply it to the KTN spacetime. We constrain the value of gravitomagnetic monopole for M87* in Sec. \ref{sec:m87}, and finally we conclude in Sec. \ref{sec:dis}. Note that the geometrized units ($G=c=1$) are adpoted throughout the paper.

\section{Kerr-Taub-NUT Spacetime}
\label{sec:1}
Before going into detail, we briefly describe the KTN spacetime below. The metric of the KTN spacetime is expressed as \cite{ml} \footnote{Here, we use the same form of KTN metric that is considered in Eqs. (1-5) of \cite{ml} and valid for $C=0$. The case $C = 0$ is the only possibility for the NUT
solutions to have a finite total angular momentum, as shown in \cite{mr}. In fact, the north and south poles play a symmetrical role for $C=0$ (see \cite{bun,vir}). However, one may repeat the same analysis (which is presented in this manuscript) for the KTN metric including $C$ (see Eqs. 4.1--4.4 of \cite{vir}). The discussions on the parameter $C$ could be found in \cite{kag,mr}.}
\begin{eqnarray} \nonumber
ds^2&=&-\f{\d}{p^2}(dt-A d\phi)^2+\f{p^2}{\d}dr^2+p^2 d\th^2 \nonumber \\
&+&\f{1}{p^2}\sin^2\th(adt-Bd\phi)^2
\label{metric}
\end{eqnarray}
with
\begin{eqnarray}\nonumber
\d&=&r^2-2Mr+a^2-n^2, \,\,\,\,\,\,\,\,\,  p^2=r^2+(n+a\cos\th)^2,
\\
A&=&a \sin^2\th-2n\cos\th, \,\,\,\,\,\,\,\,  B=r^2+a^2+n^2
\end{eqnarray}
where $M$ is the mass, $a_*=a/M$ is the Kerr parameter or spin parameter and $n_*=n/M$ is the NUT parameter or gravitomagnetic monopole of the collapsed object. The boundaries of the outer horizon is located at
\begin{eqnarray}
 r_h &=& M(1+\sqrt{1+n_*^2-a_*^2}) .
\end{eqnarray}
Setting $p^2=0$, one can obtain the location of singularity \cite{mcd} at
\begin{eqnarray}
r=0 \,\,\,\,\, {\rm and} \,\,\,\, \th_s = \cos^{-1}(-n_*/a_*),
\label{sing}
\end{eqnarray}
in KTN spacetime. The above expression (Eq. \ref{sing}) reveals that the singularity does not arise for $|n_*| > |a_*|$, which indicates a singularity-free KTN BH, whereas for a KTN BH with $n_* = a_*$, singularity arises at $\th_s=\pi$, covered by the horizon. The singularity can arise for $|n_*| \leqslant |a_*|$, which could be a KTN BH or a KTN NS depending on the numerical values of $a_*$ and $n_*$. 
Now, as the $r_h$ vanishes for $|a_*| > \sqrt{1+n_*^2}$, one can always obtain a KTN NS in this case, whereas a KTN BH with singularity (covered by the horizon) arises if the following condition is satisfied: $|n_*| \leq |a_*| \leq \sqrt{1+n_*^2}$. 

One can see an interesting fact from Eq. (\ref{sing}) that 
the location of singularity (arisen only for $|n_*| \leqslant |a_*|$) can vary depending on the sign of $a_*$ and $n_*$. In Kerr spacetime ($n_*=0$), the singularity always lies at $\th_s=\pi/2$, whether $a_*$ represents the prograde rotation ($a_* > 0$) or the retrograde rotation ($a_* < 0$). However, the presence of gravitomagnetic monopole can shift the location of singularity in the KTN spacetime, as seen from Eq. (\ref{sing}). For instance, four possibilities can be arisen, which is discussed below dividing into four quadrants.

Quadrant I ($a_* > 0$ and $n_* > 0$): The location of singularity can vary from $\th_s \rightarrow \pi/2$  to $\th_s =\pi$ for $n_* \rightarrow 0$ to $n_*=a_*$, or, for $a_* \rightarrow \infty$ to $a_*=n_*$. 

Quadrant II ($a_* < 0$ and $n_* > 0$): The location of singularity  can vary from $\th_s \rightarrow \pi/2$  to $\th_s =0$ for $n_* \rightarrow 0$ to $n_*=-a_*$, or, for $a_* \rightarrow -\infty$ to $-a_*=n_*$.  

Quadrant III ($a_* < 0$ and $n_* < 0$): The location of singularity can vary from $\th_s \rightarrow \pi/2$  to $\th_s =\pi$ for $n_* \rightarrow 0$ to $-n_*=-a_*$, or, for $a_* \rightarrow -\infty$ to $-a_*=-n_*$.  

Quadrant IV ($a_* > 0$ and $n_* < 0$): The location of singularity can vary from $\th_s \rightarrow \pi/2$  to $\th_s =0$ for $n_* \rightarrow 0$ to $-n_*=a_*$, or, for $a_* \rightarrow \infty$ to $a_*=-n_*$. 

We should note here that all the above mentioned four quadrants also include the singularity-free KTN BH regions which arises due to $|n_*| > |a_*|$.

\section{Shadow of a collapsed object}
\label{sec:2}

There are three types of photon trajectories fired from the large radii around
a collapsed object: capture, scatter to infinity and critical curve. The latter
or unstable curve separates the first and second types. If the 3-momentum of a light ray is nearly tangential to the circular photon orbit, the orbit is unstable
and this ray orbits around the collapsed object several times. This light ray is
either scattered to infinity or captured by the central object due to a small
perturbation. As they orbit around the collapsed object several times, they create a brighter region around a central dark region in the sky plane of
distant observer. This can be projected as a 2D image and the dark region of it
is called as the shadow of that collapsed object.

In order to study shadow of a collapsed object, one needs to solve the geodesic
equations. Here, we solve the geodesic equations numerically for the KTN metric
using the ray tracing code. This is an initial value problem and to calculate
the boundary of shadow we use the simulation {\color{black}with the assumption of stationarity} \footnote{{\color{black}It is important to mention here that the assumption of stationarity removes the
possibility of temporal signatures in the simulation. Thus, in practice a trajectory of the light exactly passing the pole where the discontinuity in $t$ (as discussed in Sec. \ref{intro}) arises, does not exist in the simulation.}}, similarly as that done in \cite{myshadow1,myshadow2}. A class of Runge-Kutta-Nystrom method is used,
which is explained in \cite{Lund:2009zzb}. We use adaptive step sizes in our
calculations with error control. We start from the observer's sky plane, the location of which is expressed in the Cartesian coordinates 
$\left( x' , y', z' \right)$ and the collapsed object is located at $\left( x , y, z \right)$. The observer's plane is located at a distance $D$ away from the collapsed object, and at an inclination angle $i$ (see the geometry of the system in Figure~1 of \cite{js2}). Photon's initial 4-momentum, ${\bf k_0}=(k_0^t,k_0^r,k_0^\theta,k_0^\phi)$, is perpendicular to the
observer plane. The image plane contains a grid and the photons are fired from
every point of the grid. The ray is traced-back from each pixel of image with
the initial condition to the collapsed object.

Now, we can write the initial condition \cite{coba} for our simulation as:
\be
\label{eq-1}
t_0 &=& 0 \, , \nonumber\\
r_0 &=& \sqrt{x'^2_0 + y'^2_0 + D^2} \, , \nonumber\\
\theta_0 &=& \arccos \frac{y'_0 \sin i + D \cos i}{\sqrt{x'^2_0 + y'^2_0 + D^2}} \, , \nonumber\\
\phi_0 &=& \arctan \frac{x'_0}{D \sin i - y'_0 \cos i} \, .
\ee
The initial condition for the photon 4-momentum is :
\be
\label{eq-2}
k^r_0 &=& - \frac{D}{\sqrt{x'^2_0 + y'^2_0 + D^2}} |k_0| \, , \nonumber\\
k^\theta_0 &=& \frac{\cos i - D \frac{y'_0 \sin i + D 
\cos i}{x'^2_0 + y'^2_0 + D^2}}{\sqrt{x'^2_0 + (D \sin i - y'_0 \cos i)^2}} |k_0| \, , \nonumber\\
k^\phi_0 &=& \frac{x'_0 \sin i}{x'^2_0 + (D \sin i - y'_0 \cos i)^2}|k_0| \, , \nonumber\\
k^t_0 &=& \sqrt{\left(k^r_0\right)^2 + r^2_0  \left(k^\theta_0\right)^2
+ r_0^2 \sin^2\theta_0  (k^\phi_0)^2} \, 
\ee
where $k^t_0$ is obtained from the condition $g_{\mu\nu} k^{\mu} k^{\nu} = 0$
with the metric tensor of flat space time.

Boundary of the shadow of a collapsed object is a closed curve in the observer's plane and it separates captured photon from scattered ones. To have image of the shadow boundary, we first define the center of a shadow $(x'_{\rm cs}, y'_{\rm cs})$ in analogy with 
the center of mass calculation,
\be
x'_{\rm cs} &=& \frac{\int \int \rho(x',y') x' dx' dy'}{\int \int \rho(x',y') dx' dy'} \,  \nonumber\\
y'_{\rm cs} &=& \frac{\int \int \rho(x',y') y' dx' dy'}{\int \int \rho(x',y') dx' dy'} \, ,
\ee
where $\rho(x',y') = 1$ represents inside of the shadow and $\rho(x',y') = 0$ represents the outside of it. \\
We consider $x'$ axis as the symmetry axis of shadow boundary. We start from the shorter segment in $x'$ axis with $\phi = 0$, and define $R \left( \phi \right)$ as the distance between each point of the boundary and the center. 

\begin{figure*}
%\begin{figure*}
  \begin{center}
  %\resizebox{0.5\textwidth}{!}{
  \includegraphics[width=6in]{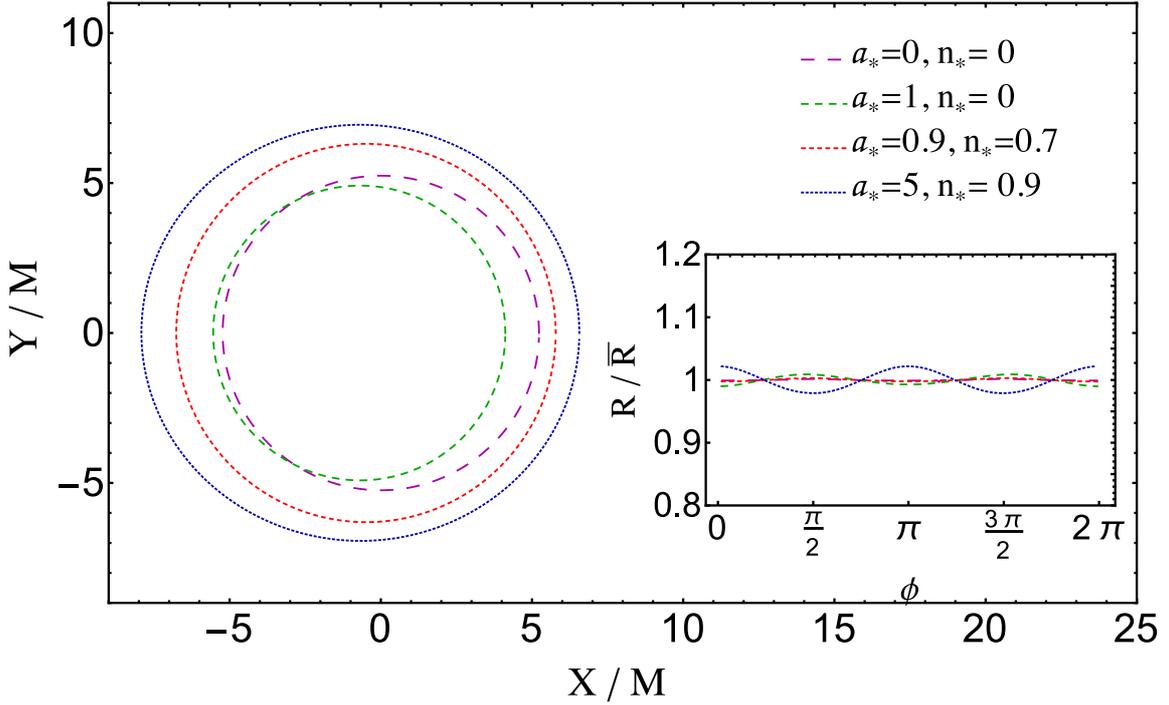}
  %  %
  \end{center}
\caption{\label{fig:shape} Shadow shapes resulting from the four different settings on the spin ($a_*$) and the NUT parameter ($n_*$), i.e., $(a_*, n_*)=(0,0)$ (Schwarzschild metric), $(1, 0)$ (extremely rotating Kerr metric), $(0.9, 0.7)$ (KTN BH), and $(5, 0.9)$ (KTN NS), respectively, for  $i=17^{\circ}$. This Figure illustrates that the shadow shapes resulting from different KTN parameter settings are all nearly circular but the shadow sizes can be significantly different from each other. In the inset, although we display that the resulting deviation from circularity is only within $5\%$ for these particular settings of parameters, it can be, in fact, higher for the different settings of $a_*$ and $n_*$.}
\end{figure*}

Figure~\ref{fig:shape} shows the examples for the cases with
$(a_*, n_*)=(0, 0)$, $(1, 0)$, $(0.9, 0.7)$, and $(5, 0.9)$, i.e., the Schwarzschild metric, Kerr metric, KTN BH, and KTN NS for a particular inclination angle, say, $i=17^{\circ}$. 
For different settings of $n_*$, as it is shown in this Figure, the shadow sizes differ significantly, but the shapes are nearly circular and only have slight differences. For instance, the shadow is circular for the Schwarzschild BH, whereas it slightly deviates from the circularity even if $a_*$ increases to a large value for the Kerr BH, as shown in \cite{EHT5}. For a fixed $a_*$, the shadow size increases with increasing $n_*$, but it is almost circular for the KTN BH cases and it becomes asymmetric for the KTN NS cases. However, the deviation from circularity is on the order of percentage level or less. For example, it is only $1.5\%$ if $(a_*, n_*)=(5, 0.9)$.

Now, it is asserted by the EHT collaboration that the recently released image of the shadow of M87* is not exactly circular \cite{EHT1}, i.e., it deviates from circularity. Therefore, one needs to define the average radius of the shadow which can be expressed as (see Eq. 4 of \cite{sc})
\be
\bar R^2 \equiv \frac{1}{2\pi} \int_0^{2\pi} \! R^2(\phi) \, \mathrm{d}\phi.
\ee
Following \cite{sc,EHT1} one may also define a parameter $\Delta d$ to describe the asymmetry of shadow by using the difference between the RMS distance and the average radius of the shadow $\bar R$ :
\be
\label{circu}
\Delta d \equiv \frac{1}{ \bar R} \,\,\sqrt{\frac{1}{2\pi} \int_0^{2\pi} \! (R(\phi) - \bar R)^2 \, \mathrm{d}\phi}\,.
\ee
Here, the asymmetry parameter $\Delta d$ quantifies the deviation from circularity of the shadow. This parameter can also be used to compare the theoretically predicted shadow size of a collapsed object with the observational one. Below we first briefly discuss the theoretically predicted values of shadow size and circularity for the KTN spacetime in Secs. \ref{sec:kerr} and \ref{sec:ktn}, respectively. Later, using the theoretical values and the recently reported observational parameter values of EHT, we constrain the gravitomagnetic monopole/NUT charge for M87* in Sec. \ref{sec:m87}.

% \section{Results}
%
\subsection{Dependence of the shadow size on $a_*$, $n_*$, and $i$}
\label{sec:kerr}

\begin{figure*}
\begin{center}
  \subfigure[$i=1^{\circ}$]{\includegraphics[width=2.9in,angle=0]{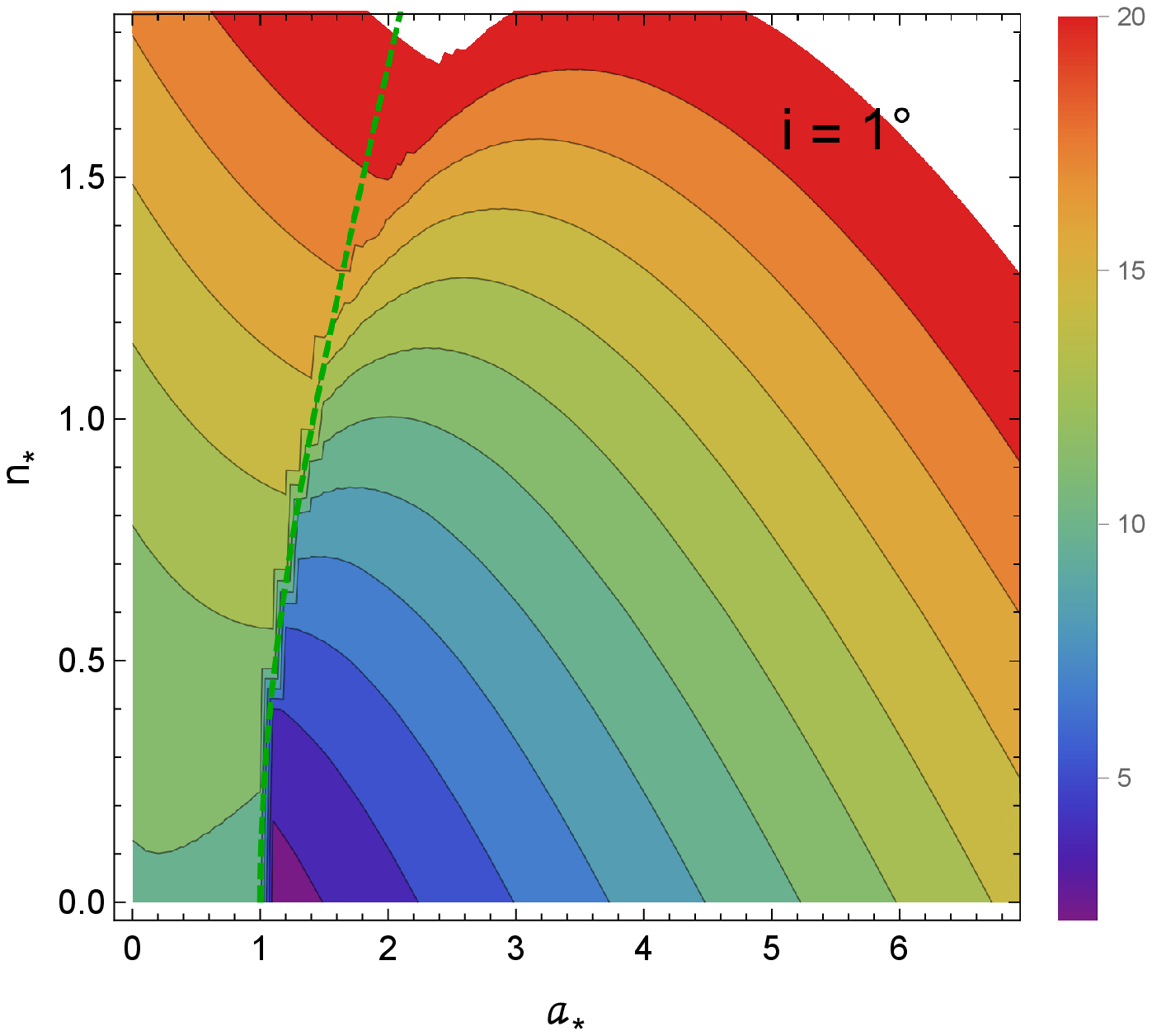}}
\hspace{0.05\textwidth}
\subfigure[$i=17^{\circ}$]{\includegraphics[width=2.9in,angle=0]{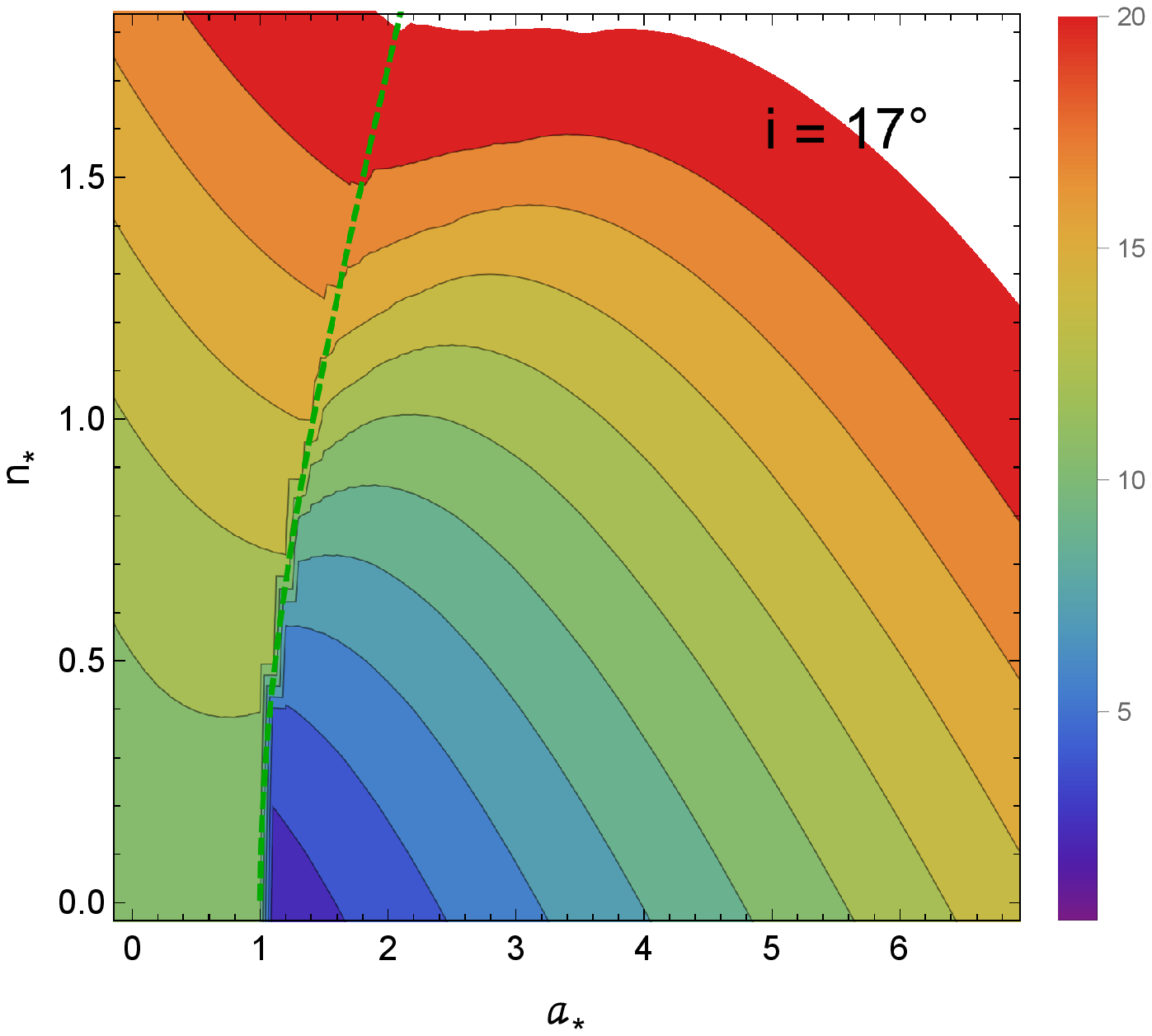}}
\hspace{0.05\textwidth}
\subfigure[$i=40^{\circ}$]{\includegraphics[width=2.9in,angle=0]{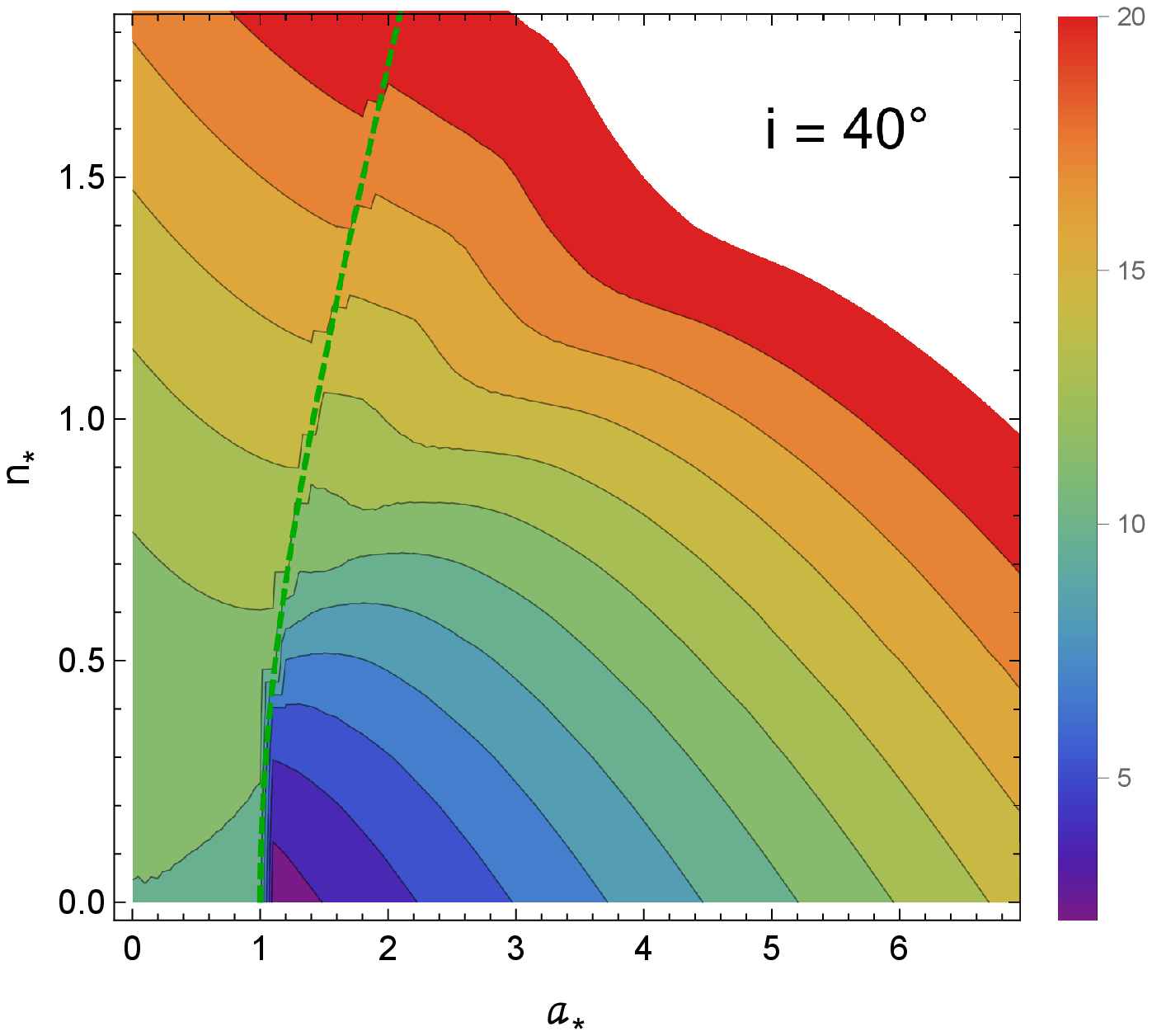}}
\hspace{0.05\textwidth}
\subfigure[$i=89^{\circ}$]{\includegraphics[width=2.9in,angle=0]{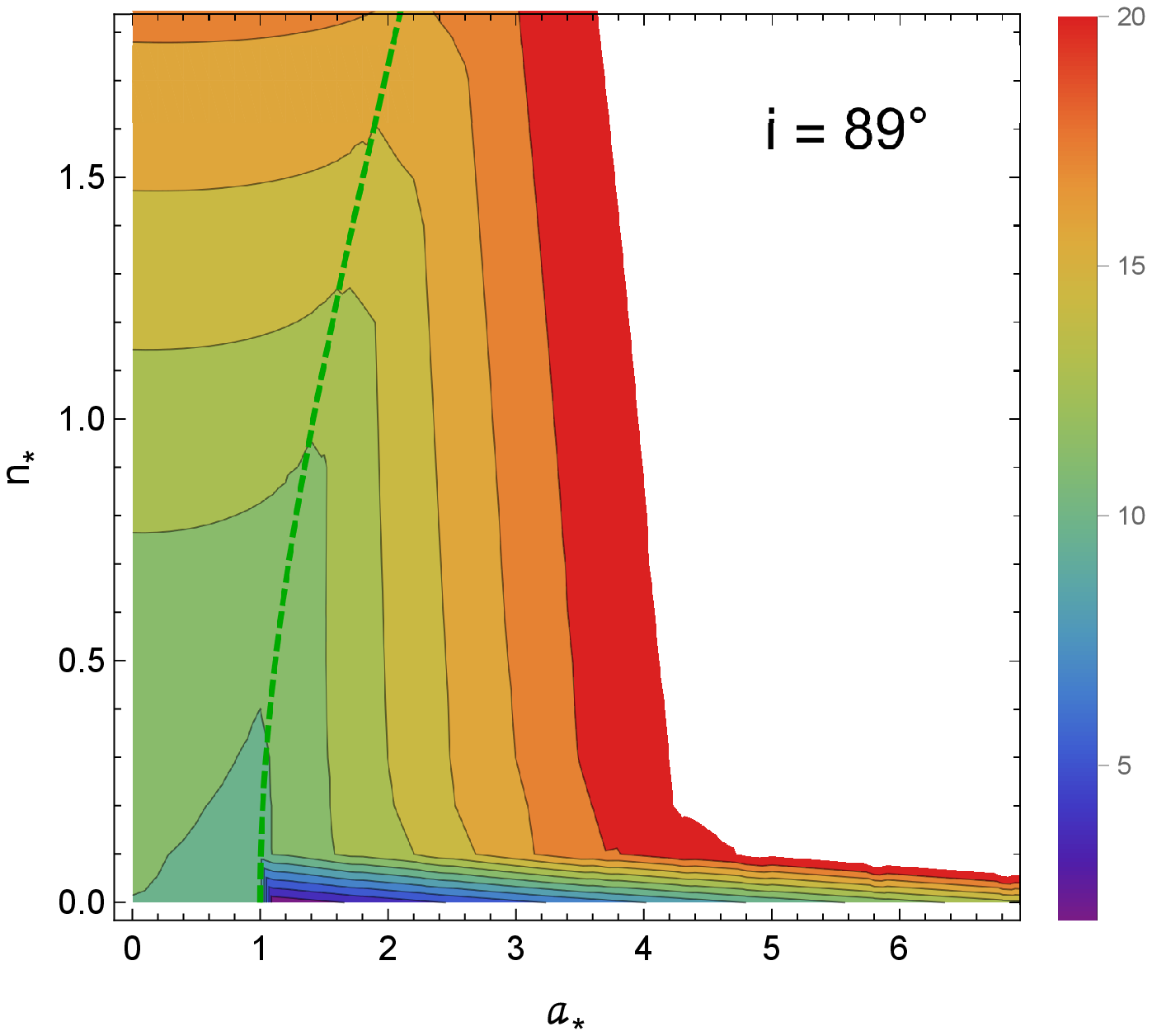}}
\caption{\label{fig:th} Dependence of the shadow radius $\bar{R}$  (in unit of $`M$')
on the Kerr and NUT parameters of a KTN collapsed object, for four different
inclination angles $(i)$. Variation of these theoretically predicted shadow sizes are
indicated by the rainbow colors. The green dashed line represents the
division between the KTN BHs (left) and the KTN NSs (right).  As seen from this
figure, compared to the shadow sizes of the Kerr BHs, the shadow sizes of the
Kerr NSs are not necessarily smaller in the whole range of $a_*$ (i.e.,
$|a_*| > 1$), and they can be larger. See Secs. \ref{sec:kerr} and \ref{sec:ktn} for details.}
\end{center}   
\end{figure*}
Figure~\ref{fig:th} is drawn for four different inclination angles to show the
variation of shadow sizes with the Kerr and NUT parameters (in unit of the mass scale $M$), as obtained
theoretically.  Our results are consistent with the shadow sizes obtained in
\cite{tak} (see also \cite{js1,js2}) for Kerr BHs.  In case of the Kerr BH, the
shadow size first decreases with increasing the value of $a_*$, until $a_*=1$.
It faces a sudden decrement in its value for further increment in $a_*$, i.e.,
$a_* = 1+\epsilon$ where $\epsilon \rightarrow 0^+$. The latter one stands for
an extreme case of Kerr NS. Interestingly, the shadow size increases with
increasing the value of $a_*$ for Kerr NS, i.e., $a_*=1+\epsilon$ to a higher
value, which is true for all inclination angles (see Figure~\ref{fig:th}). The
physical reason behind this may be understood by considering the example of
the behavior of corotating equatorial circular photon orbit (CPO).  In the
case of the Kerr BH, the radius of CPO, $r_{\rm CPO}$, comes closer and closer
to the event horizon with increasing the value of $a_*$ (i.e., $3 \geq r_{\rm
CPO}/M \geq 1$ for $0 \leq a_* \leq 1$), whereas $r_{\rm CPO}$ is always
located (formally) at the ring singularity ($r=0, ~ \th=\pi/2$) \cite{chastu}
in the case of the Kerr NS.  This means that the $r_{\rm CPO}$ can exist very
close to the ring singularity for all values of $a_* : a_* > 1$, in principle.
One intriguing behavior of the ring singularity that emerges is, its radius
increases with increasing $a_*$ (see Figure 2 of \cite{ckj}), and
hence, the CPO also becomes bigger and bigger in principle. This can be
realized from the Kerr-Schild coordinates which reduce to \cite{ch} :
\begin{eqnarray}
x^2+y^2=(r^2+a^2)\sin^2\th \,\,\,\, {\rm and} \,\,\,\, z=r\cos\th .
\label{ks}
\end{eqnarray}
Thus, the ring singularity ($r=0, ~ \th=\pi/2$ in the Boyer-Lindquist coordinates) can be expressed as (see Figure 25 of \cite{ch})
\begin{eqnarray}
x^2+y^2=a^2 \,\,\,\, {\rm and} \,\,\,\, z=0.
\label{rs}
\end{eqnarray}
Eq. (\ref{ks}) shows that the radius (in the Kerr-Schild
coordinates) of $r_{\rm CPO}$ ($\equiv M$) for $a_*=1$ is $\sqrt{x^2+y^2}|_{r
\rightarrow M}=\sqrt{2}M=1.414M$. Now, for a nearly extremal naked singularity (NENS), say,
$a_*=1.00001$, the radius of CPO is : $\sqrt{x^2+y^2}|_{r \rightarrow
0}=|a_*|=1.00001M$. It is needless to say here that the  radius of CPO
increases with the further increment of $a_*$ for the Kerr NS. The above
discussion gives a rough idea that why the shadow size first decreases for the
`transition' from the extremal BH to a NENS and then increases
again.

\begin{figure*}
\begin{center}
  \includegraphics[width=6.5in,height=5.2in]{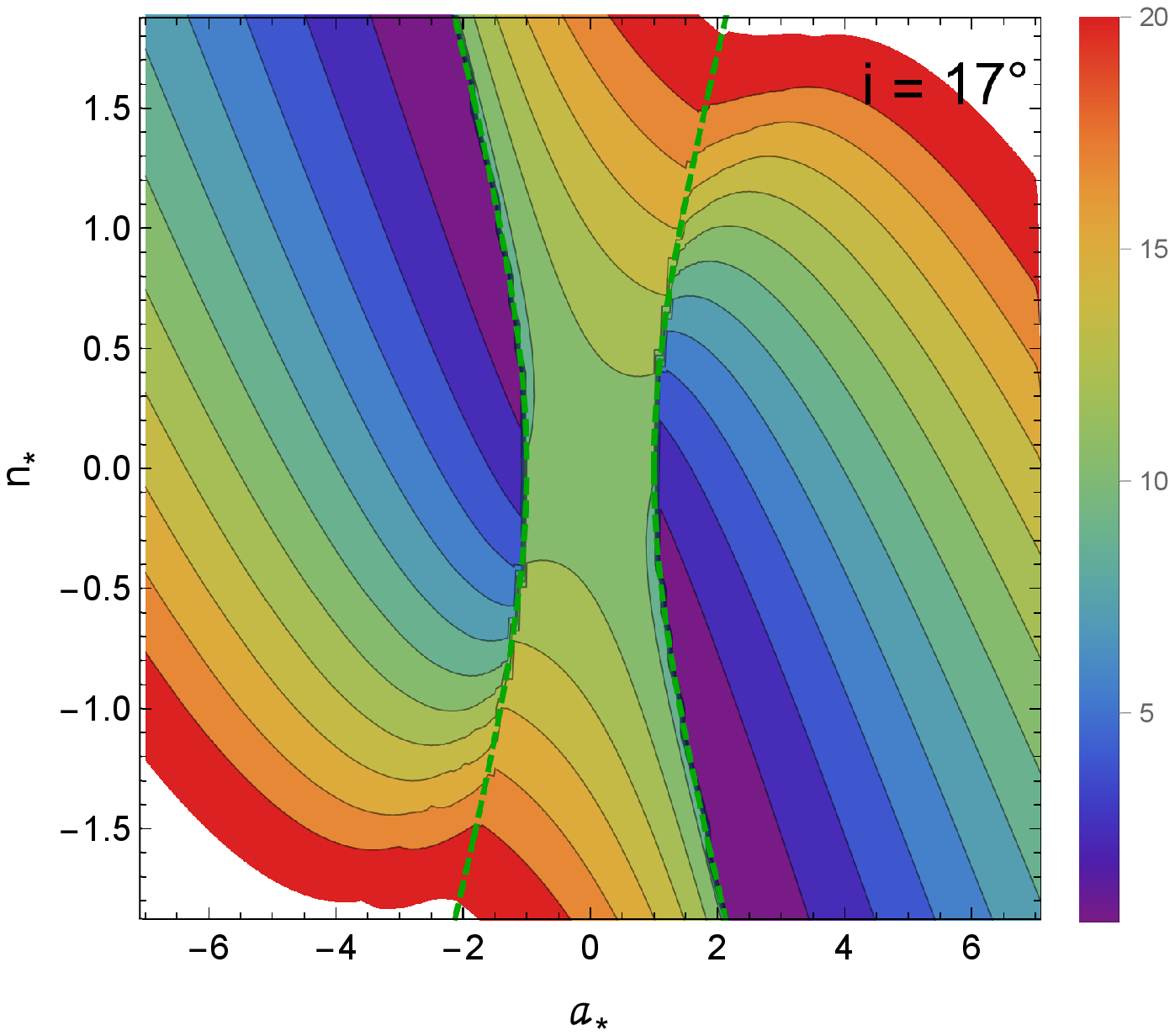}
\caption{\label{nai17_4q} Dependence of the shadow radius $\bar{R}$  (in unit of $`M$')
on the Kerr and NUT parameters of a KTN collapsed object, for $i=17^{\circ}$, i.e., it is an extended version of Panel (b) of Figure  \ref{fig:th} with all of the four quadrants: Quadrants I--IV. The plane is divided into the KTN BH and KTN NS regions by the two thin dashed green lines, i.e., the region between the two green lines implies the KTN BH region whereas rest of the plane implies
the KTN NS region. This figure shows that the shadow size increases with increasing $|n_*|$ for a fixed value of $|a_*|$ in case of the KTN BH. On the other hand, it increases (or decreases) with increasing $n_*$ for a fixed $a_* > 0$ (or $a_* < 0$) in the range of  $n_* > -\cot 17^{\circ}$ (or $n_* < \cot 17^{\circ}$)  for KTN NS. Here we plot for a limited parameter space. For a general large parameter space ($|n_*| > \cot 17^{\circ}$), the violet colored region adjacent to the green dashed line, shifts towards the right (left) side of the plot in Quadrant IV (Quadrant II), and the subsequent rainbow color appears at adjacent to the green dashed line. This means that the shadow sizes of KTN NENSs can be much bigger ($\sim 10$) than zero in such a special case. See Sec. \ref{sec:kerr} and Appendix \ref{app} for details.}
\end{center}
\end{figure*}

Note here that the shadow size varies from $10.4$ to $9.6$ for the Kerr BH with
spin ($a_*$) varying from $0$ to $1$ for
the inclination angles, i.e., $0 \leq i \leq 89^{\circ}$ (see Figure ~2 of \cite{tak}). One can see exactly the opposite scene for the Taub-NUT BH, i.e., KTN BH with
$a_* = 0$. In this case, the shadow size increases with increasing the value of
$|n_*|$, but it does not change with the value of $i$, as the Taub-NUT spacetime
is spherically symmetric \cite{mis,lnbl}. The different nature of these two
parameters is reflected in the shadow size of KTN BH.
In general, the shadow size increases with increasing $|n_*|$ for a fixed value of $a_*$ in case of the KTN BH. In contrast, it can increase or decrease with increasing $n_*$ for a fixed value of $a_*$ in case of the KTN NS.
This statement is true for any inclination
$i$ no matter whether it is a KTN BH or a KTN NS, which is also clear from Panels (a)--(d) of Figure~\ref{fig:th} as well as \footnote{Figure  \ref{nai17_4q} is, in fact, the extensive version of Panel (b) of Figure~\ref{fig:th} with all of the four quadrants, as this particular plot is also necessary to constrain $n_*$ and $a_*$ for M87* using Figure \ref{fig:na}. As the feature of the shadows for all inclination angles are almost similar to Figure \ref{nai17_4q}, we do not repeat it by plotting Panels (a), (c) and (d) of  Figure~\ref{fig:th} for all of the four quadrants.}Figure \ref{nai17_4q}.

\begin{figure*}
\begin{center}
  \subfigure[$a_*=1.1$]{\includegraphics[width=2.9in,angle=0]{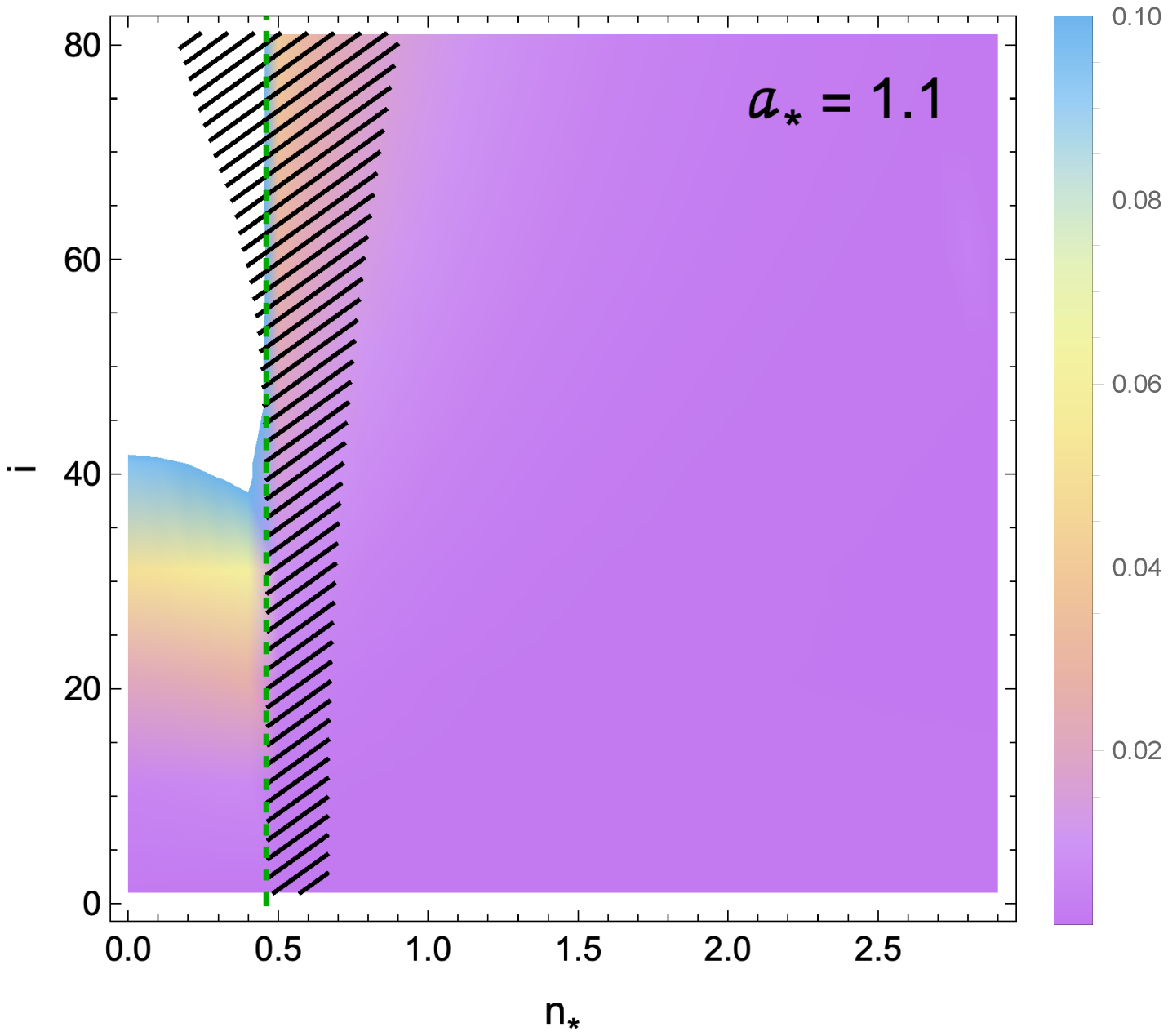}}
  \hspace{0.05\textwidth}
\subfigure[$a_*=2.5$]{\includegraphics[width=2.9in,angle=0]{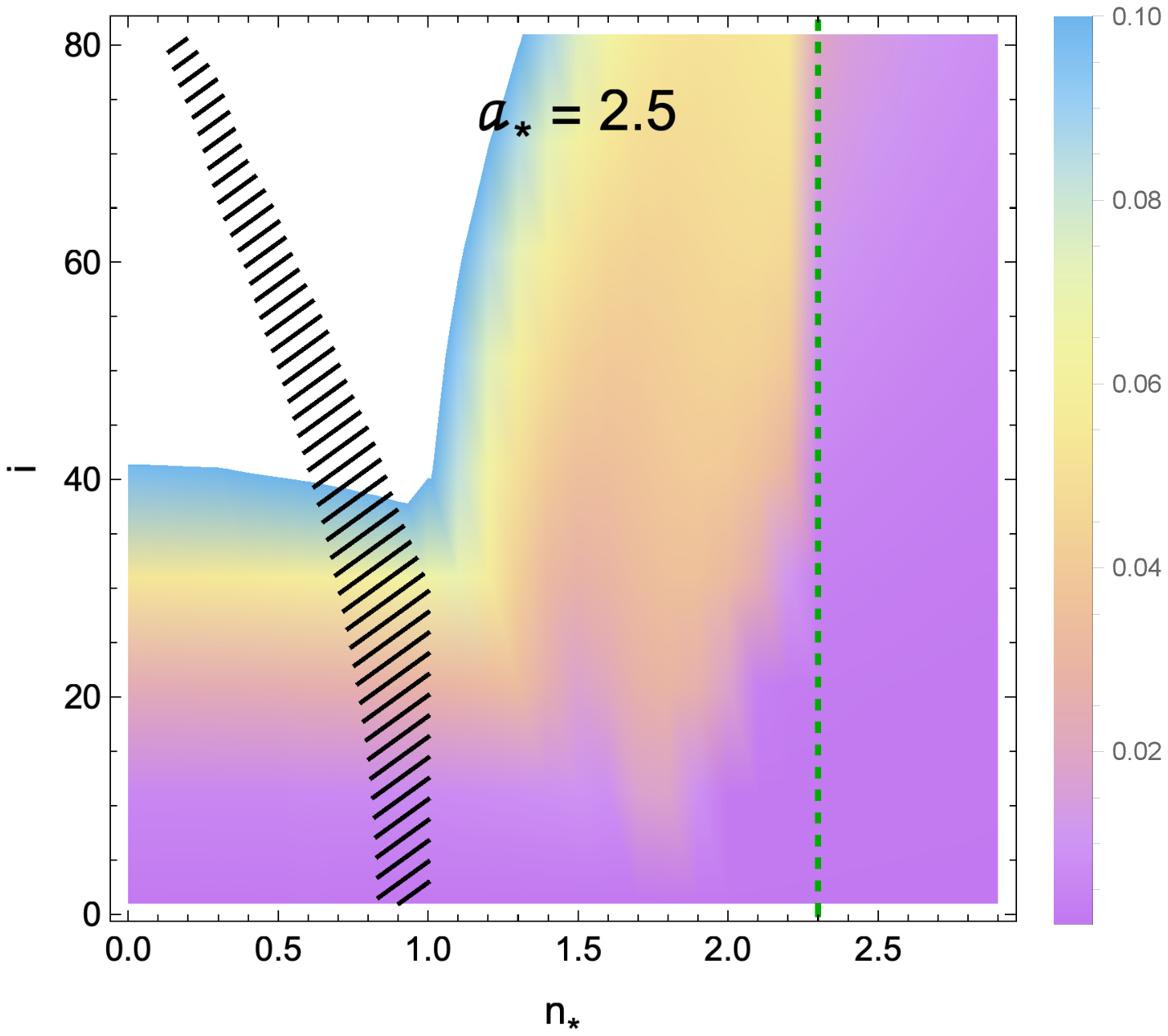}}
\caption{\label{fig:in} Dependence of the deviation from the circularity of an object with
KTN metric on $i$ and $n_*$ plane, for two different $a_*$ values. These two panels illustrate one
important property of the NUT parameter in general, i.e, it decreases the deviation from
the circularity of the BH shadow.
In each panel, the hatched region indicates the parameter space allowed by the
shadow size measured from the EHT observations for M87* (Eq.~\ref{size}),
while the white region is excluded by the circularity constraint from the same EHT
observation ($\Delta d \lesssim 0.1$).  The left side of the thin green dashed line
stands for the KTN NSs, and its right side the KTN BHs. See Secs. \ref{sec:ktn} and \ref{sec:m87} for details.}
\end{center}
\end{figure*}

The appearance of the special feature (decreasing the shadow size for the transition from the extremal BH to the NENS) in the violet colored region of Figure \ref{nai17_4q} is not unusual in case of a NENS as discussed at the end of the first paragraph of this section. However, this explanation is true only for the KTN spacetime with the value of $n_* : 0 > n_* > -3.27$, if one moves along the $OY'$ axis with a constant value of $a_*$. For a general large parameter space, the violet colored regions adjacent to the green dashed lines, shifts towards the right (left) side of the plot in Quadrant IV (Quadrant II), and the subsequent rainbow color appears at adjacent to the green dashed line. This means that the shadow sizes of KTN NENSs can be much bigger ($\sim 10$) than zero in such a case. Not only this, it is also shown that if $\th_s$ (location of the singularity, see Eq. \ref{sing}) is bigger than the inclination angle ($i$), i.e., $\th_s > i$, one can see an extremely smaller shadow, whereas the shadow will be much bigger for $\th_s \leqslant i$. The above statement is true in either way, i.e., depending on a slight change  in $i$ from $\th_s$ (i.e., $i \rightarrow \th_s \pm 0.1^{\circ}$), one can see a large difference in the shadow size of a same KTN object. For example, if we consider a KTN NS with $(a_*, n_*) \sim (8.36, -8)$, i.e., $\th_s \sim 17^{\circ}$, one can see the shadow size of the object as $\sim 0.003$ with $\Delta d \sim 0$ for $i =16.9^{\circ}$, whereas the shadow size becomes $\sim 15.5$ with $\Delta d \sim 0.09$ for $i=17.1^{\circ}$. The reason behind these interesting discontinuity effects is due to an important relation between $i$ and $\th_s$, which is vividly discussed in Appendix \ref{app}.

\subsection{Dependence of the shadow circularity on $a_*$,
$n_*$, and $i$}\label{sec:ktn}
 
Earlier it was shown in \cite{abdu} that the NUT parameter not only increases the size of the KTN BH
shadow (with the same $M$ and $a_*$), but it also circularize the BH shadow
ellipse{\footnote{Although it is asserted as the `ellipse' in \cite{abdu}, it
should be actually regarded as the `distorted form of a circle'. The increment
of BH's angular momentum distorts the `circular' form of shadow but it does not
take the shape of shadow to an exact form of `ellipse'. See Figure 4 and the
end of Sec. 3 of \cite{abdu} for details. However, this simpler description
helps one to realize here the `opposite behavior' of the Kerr parameter and
the NUT parameter in a better way. Therefore, we continue our discussion in
this paper following \cite{abdu}, i.e., we use the word `ellipse' or
`elliptical', instead of the `distorted form of a circle', keeping in mind that
the word `ellipse' does not mean here an exact form of ellipse or elliptical
shape in mathematics.}} \cite{abdu}. The first property can be seen from
Figure~\ref{fig:th}, and the second one from Figure~\ref{fig:in}. The left side
of green dashed line of Figure~\ref{fig:in} represents the KTN NS region
whereas its right side represents the KTN BH region. Panel (a) of
Figure~\ref{fig:in} shows that the shadow of KTN BH becomes circular with
increasing the value  of $n_*$ for $a_*=1.1$.  The Kerr parameter
tries to make it further elliptical.  Comparing panels (a) and (b) of
Figure~\ref{fig:in}, one can see this interesting difference of the Kerr and
NUT parameters for the KTN BH. The same figures also show that the shadows are
in general elliptical for $i \sim 40^{\circ}$ in the case of KTN NSs.

\begin{figure*}
\begin{center}
\subfigure[$n_*=0.1$]{\includegraphics[width=2.9in]{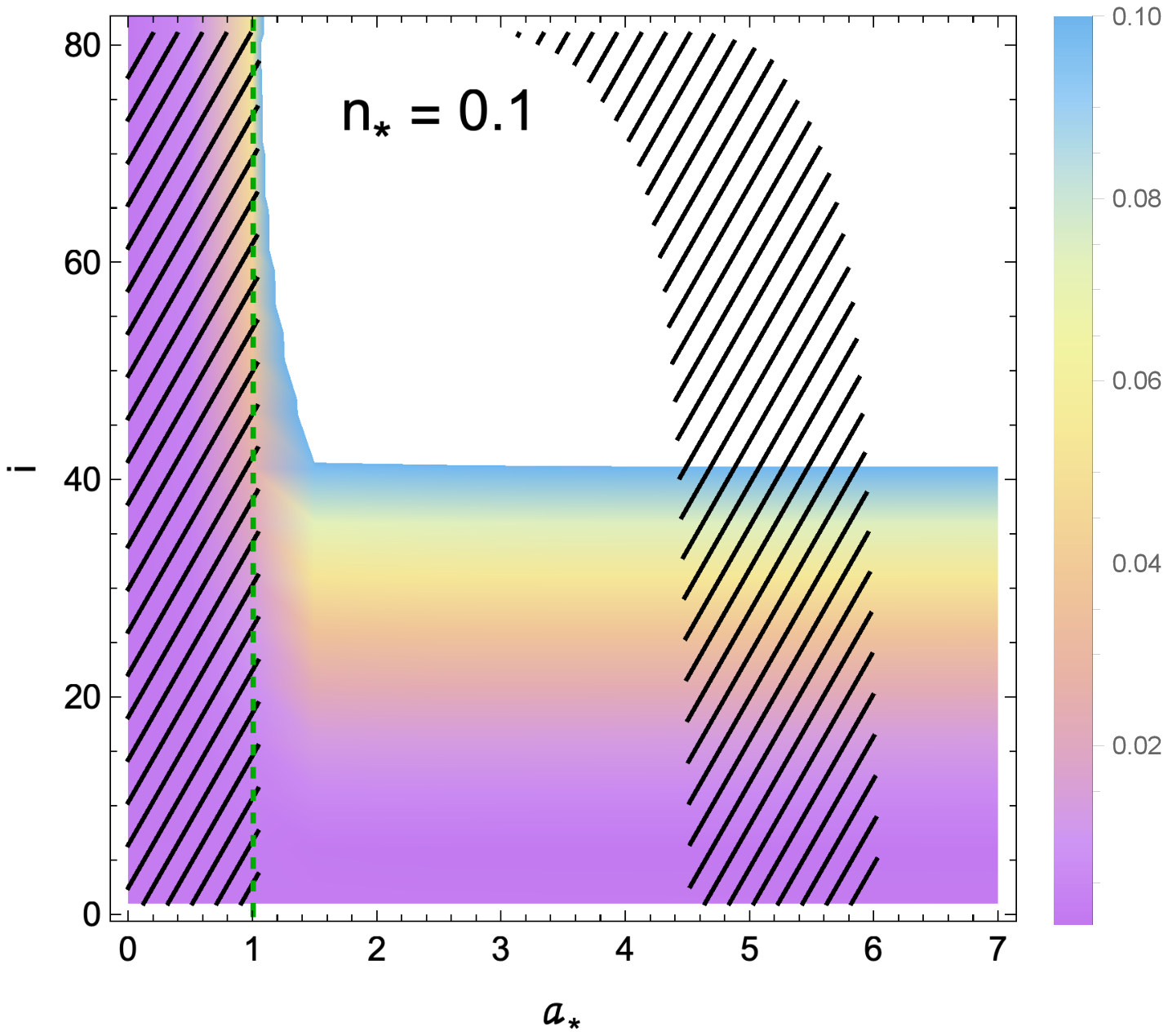}}
  \hspace{0.05\textwidth}
 \subfigure[$n_*=0.3$]{\includegraphics[width=2.9in,angle=0]{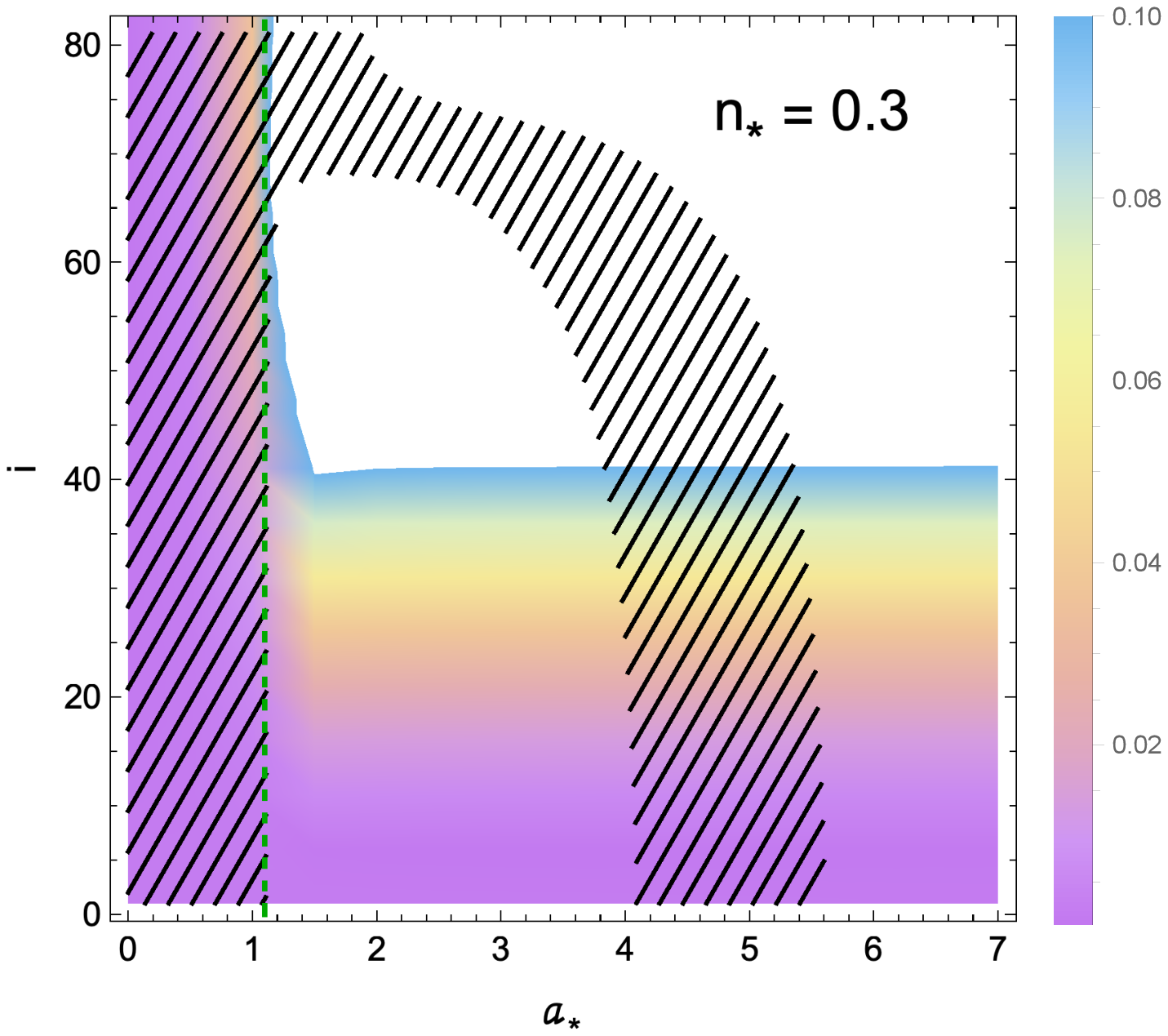}}
\subfigure[$n_*=1.1$]{\includegraphics[width=3.in]{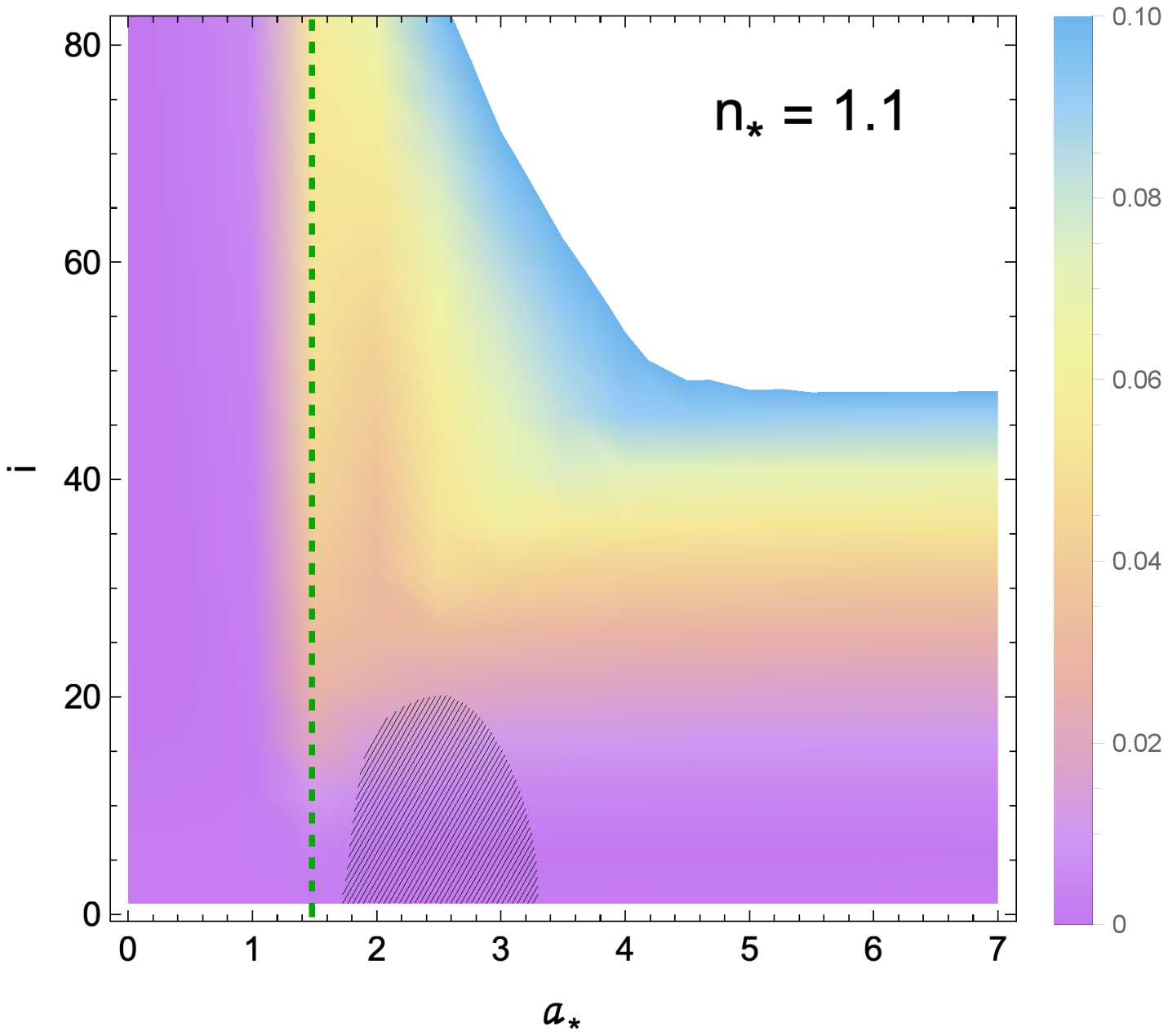}}
 \hspace{0.05\textwidth}
\caption{\label{fig:ia} Dependence of the deviation from the circularity of an
object with KTN metric on $i$ and $a_*$, for the different $n_*$
values. In each panel, the hatched region indicates the parameter space allowed
by the EHT measurement of M87* shadow size, and the white region is excluded by
the circularity constraint obtained for M87* ($\Delta d\lesssim 0.1$). The
green dashed line stands for the extremal KTN BH. The left and the right sides of
green dashed line represent the KTN BH and the KTN NS regions, respectively. See
Secs. \ref{sec:ktn} and \ref{sec:m87} for details..} 
\end{center}  
\end{figure*}
Now, due to the small increment in $n_*$ from $0$ (i.e., equivalent to the Kerr spacetime) to $0.1$, Panel (a) of Figure~\ref{fig:ia} spans the inclination and spin
parameter region. The colored region represents the  deviation from
circularity which is less than $10\%$ or $0.1$ as defined in
Eq.~(\ref{circu}). As we have already mentioned that our probe includes both the
BH as well as NS region, the left side of the green dashed line represents the
KTN BHs whereas right side of the same represents the KTN NSs. The green dashed
line stands for the extremal KTN BH. The maximum allowed value of inclination
for $a_* \sim 6$ is less than $40^\circ$.  If we further
increase (decrease) the value of $n_*$, the shadow size will be larger
(smaller) and the hatched region in KTN NS moves to the smaller (larger) value
of the spin parameter. This can be seen by comparing Panels (a) and (b) of
Figure~\ref{fig:ia}.

Figure~\ref{fig:na} is drawn with the combinations of $a_*$ and $n_*$ for a fixed inclination angle, $i=17^{\circ}$.
The presence of thin Yellow-Cyan (YC) region along the green dashed line of this figure indicates that a high deviation of circularity with $3 \% < \Delta d \lesssim 10\% $ can be possible in Quadrants II and IV. For the general large parameter space, 
the thin YC regions, adjacent to the green dashed lines of Quadrants II \& IV, are broadened for increasing the value of $|n_*|$.

\section{\label{sec:m87}Constraining the gravitomagnetic monopole in M87*}~

It was suggested in \cite{EHT5} that the possibility for M87* being a Kerr NS
is ruled out by arguing that the shadows of Kerr NSs ($|a_*| > 1$) are
substantially smaller and very asymmetric compared to those of Kerr BHs.
Referring to the discussion of Sec. \ref{sec:kerr}, we can say that the argument given in
\cite{EHT5} is shaky. This can also be seen from Figure~\ref{fig:th} (b)
that one can obtain a short range of $a_*: 4.5 \lesssim a_* \lesssim 6.5$ for
which the shadow size is comparable to the Kerr BHs. This is recently shown by
\cite{sc} as well. However, there is a basic difference between
that work \cite{sc} and our present work. They considered one
extra parameter, $R_{\rm ss}$, which governs the scale at which quantum
gravity effects become relevant. Therefore, they obtained two different ranges
of $a_*$ for which M87* could be a superspinar. These are: 
$1 \lesssim a_* \lesssim 4.5$  
for 
$1.8 \lesssim R_{\rm ss}/M \lesssim 3.5$ 
and 
$4.5 \lesssim a_* \lesssim 6.5$ 
for 
$ R_{\rm ss}/M \lesssim 3$. 
In our case, we do not consider that extra parameter $R_{\rm ss}$ for the KTN
NS. This means, we consider up to $r \rightarrow 0$, i.e, $R_{\rm ss}
\rightarrow 0$ is considered for our work.  That is why, we obtain (see
Figure ~\ref{fig:th}(b)) the allowed range for M87* as $4.5  \lesssim a_* \lesssim
6.5$ for $n_* \rightarrow 0$.  It is useful to mention here that we restrict
our probe for the KTN NS up to $R_{\rm ss}\equiv r \rightarrow 0$ in this paper.  We do not consider the KTN NS with a boundary at a positive
value of $r$ unlike those presented in \cite{sc}.  This also means that the
quantum gravity effects are considered not to be relevant \cite{gh,ckp} for $r
> 0$ for this work. 

To quantify our results we first consider the recently reported observational constraints by the EHT collaboration \cite{EHT6} on the shadow size of the BH in M87 as \cite{EHT1,sc} 
\be
\label{size}
\frac{D\delta}{M} \simeq 11.0 \pm 1.5
\ee
for $M$ (mass of M87*) $= (6.5 \pm 0.2|_{\rm stat} \pm
0.7|_{\rm sys}) \times 10^9 M_{\odot}$ and $\delta$ (average diameter of the
crescent) $= (42 \pm 3)\, \mu$arcsec \cite{EHT6}. Although the distance of M87* is $D =
16.8^{+0.8}_{-0.7}$\,Mpc, we consider $D=16.8\pm 0.75$~Mpc
in the above equation for the simplification of our calculation following \cite{sc}. Considering the KTN metric (Eq. \ref{metric}) and observed parameter values, one can now constrain the value of NUT charge in M87* for Eq.~(\ref{size}). It is needless to say here that we use the KTN metric instead of the Kerr metric which was used in \cite{EHT1,sc}. We also follow the limit of $\Delta d$: $\Delta d \lesssim 10\%$ \cite{EHT1}, as reported by the EHT collaboration.

Here, Figure~\ref{fig:in} shows how the allowed region of
the shadow size of M87* varies with $n_*$ for a fixed value of $a_*$, irrespective of the inclination angle. Especially the hatched region of this figure shows the allowed size of shadow
according to Eq.~(\ref{size}). This implies that although the high value of inclination angle is satisfied for the KTN BH, it is excluded for the KTN NS by the circularity condition of the shadow of M87*. If we increase the
value of $a_*$, the size of the shadow will be smaller and the hatched region moves from the right to left, i.e., BH to NS region. It also shows that allowed
values of the inclination angle decreases with the increasing value of $a_*$. Interestingly, the shadow size of M87* does not only exclude the KTN BH for $n_*=1.1$, but also excludes the high value of inclination angle. This is shown by the hatched region of Panel (c) of Figure~\ref{fig:ia} that the following range of $a_*: 1.7 < a_* < 3.3$ with $i < 20^{\circ}$ is only compatible with the circularity condition of M87* for this particular case. 

\begin{figure*}{h}
\begin{center}
\includegraphics[width=7in]{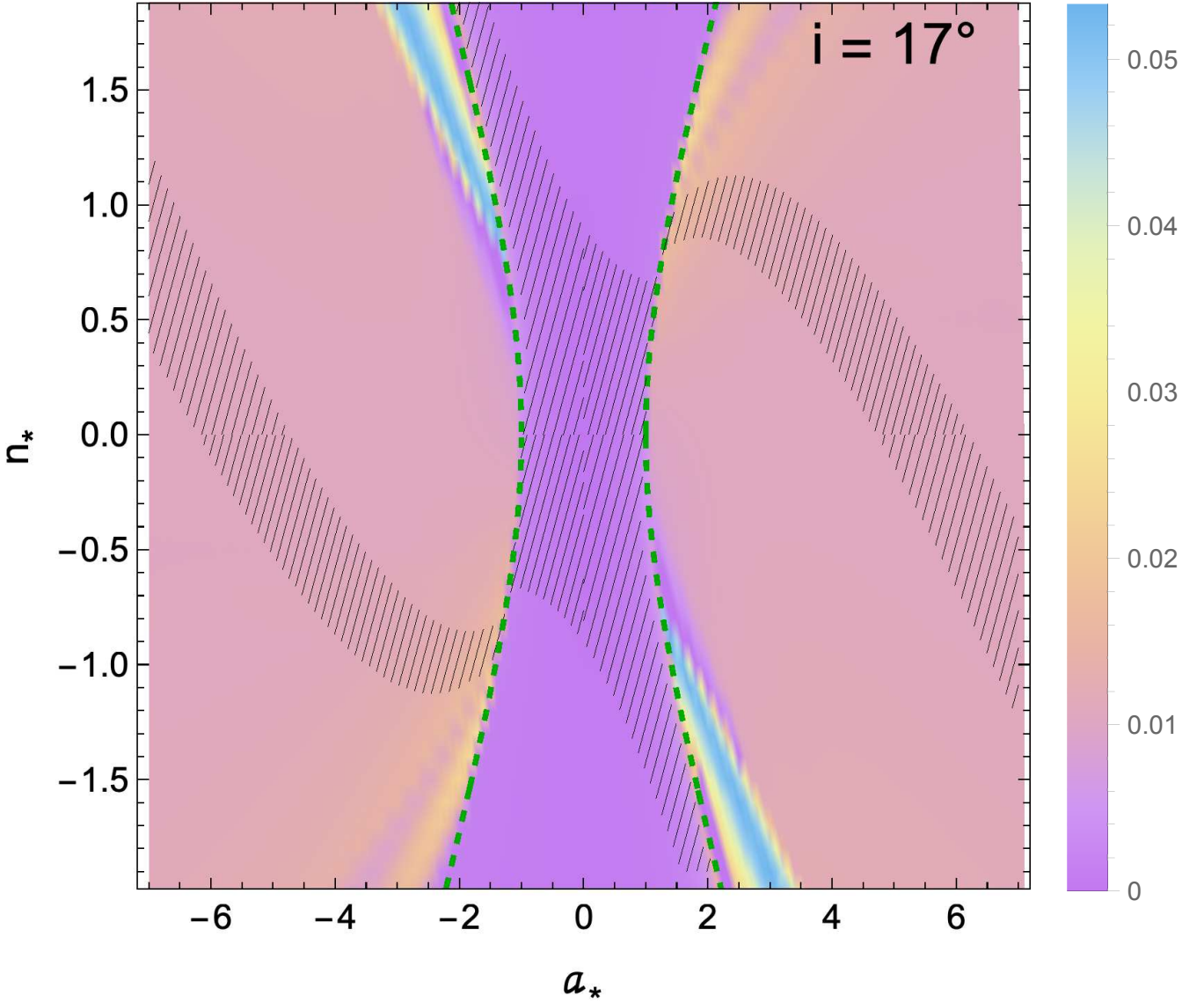}
\caption{\label{fig:na} Constraints on NUT parameter ($n_*$) and Kerr parameter
($a_*$) obtained by using the EHT results for M87* shadow and assuming an
inclination angle of $i=17^{\circ}$. The plane is divided into KTN BH and KTN
NS regions by the two thin dashed green lines, i.e., the
intermediate portion of the two green lines implies the KTN BH region whereas
rest of the plane implies the KTN NS region. The {\color{black}entire (colored) region of this plot} represents
those with circularity $\Delta d \lesssim 10\%$, as required by the EHT
observations for M87*. The combinations of $a_*$ \& $n_*$ in the hatched region can be regarded as the allowed region for M87*. This figure shows that M87*
could contain gravitomagnetic monopole $n_*$, in which case the upper limit of $n_*$ cannot be greater than $1.1$, i.e., $n_* \lesssim 1.1$ for the prograde rotation ($a_* > 0$), whereas the lower limit
of $n_*$ cannot be less than $-1.1$, i.e., $ n_* \gtrsim -1.1$ for the retrograde rotation ($a_* < 0$). This figure also suggests that the measurement of the
shadow asymmetry with 1\% of accuracy can help to break the degeneracy between
a BH and a NS, in general. Even if the KTN NS is falsifiable, M87* could still be described as a KTN BH with the upper limit of $n_*$ as $n_* < 0.9$ in case of the prograde rotation, and the lower limit of $n_*$ as $ n_* > -0.9$ in case of the retrograde rotation. Here we plot for a limited parameter space. For a general large parameter space, the thin Yellow-Cyan regions, adjacent to the green dashed lines of Quadrants II \& IV, are broadened for increasing the value of $|n_*|$ and the black hatched region of NS can be overlapped with this region. Moreover, the value of $\Delta d$ in the black hatched regions of NS is not always bounded within the value $10\%$, but it can also be equal to or greater than $10\%$ (i.e., $\Delta d \lesseqgtr 10\%$ is possible in the black hatched region of NS) depending on the values of $a_*$ \& $n_*$ in the general parameter space. See Secs. \ref{sec:ktn} and \ref{sec:m87} for details.}
\end{center}
\end{figure*}

Now, as mentioned, Figure~\ref{fig:na} is drawn for $i=17^{\circ}$ to show the combinations of $a_*$ and $n_*$ which could be relevant for M87*. This assumption is made based on the estimated angle between the approaching jet (parallel/anti-parallel to spin direction) and line of sight by \cite{jet}, that leads to the value as $17^{\circ}$. The jet direction is controlled by the BH spin. If the BH spin axis is aligned with the jet, then the asymmetry of the shadow implies that the black hole spin is pointing away from Earth, i.e., the rotation of the BH is clockwise ($a_* < 0$) as viewed from Earth (see Fig. 5 of \cite{EHT5}).
In Figure~\ref{fig:na}, we plot all of the four quadrants, as discussed at the end of Sec. \ref{sec:1} (we do not repeat here it again). A close observation of our results reveals that the features of hatched region in Quadrants I and III are same, whereas the features of Quadrants II and IV are same. Thus, we describe below these two different features separately.

\subsection{Quadrants I and III} One can see from the plots of Quadrants I and III
that M87* is not allowed as a KTN BH for the high value of $|n_*|$, i.e.,
$|n_*| > 1$. For instance, if M87* is a KTN BH with $|a_*| = 1$, the value of
$|n_*|$ has to be less than $0.6$. If M87* is a KTN BH, the allowed ranges are
$0 < |a_*| < \sqrt{1+n_*^2}$ and $0 < |n_*| < 0.9$. To be a KTN
NS, the allowed ranges are $1 < |a_*| < 6.2$ and $0 < |n_*| \lesssim 1.1$. The
upper limit of the NUT charge which could be contained in M87*, is $|n_*| \sim
1.1$. In that case, the value of the spin parameter will be $|a_*| \sim 2.4$.
Quadrants I and III of Figure~\ref{fig:na} and the above mentioned two separate
ranges imply that there is a degeneracy in the value of $|n_*|$, i.e. for $0 <
|n_*| < 0.9$, M87* could be a KTN BH or a KTN NS. This degeneracy could be
broken by analyzing the deviation of circularity of the shadow of M87* more
accurately.  To be a KTN BH with $0 < |n_*| < 0.9$, the shadow
should be almost circular (i.e., $\Delta d \lesssim 1\% $) whereas, it can significantly deviate from circularity if M87* is a KTN NS with $0 < |n_*| < 0.9$, as seen from
Quadrants I and III of Figure~\ref{fig:na}.

\subsection{\label{q24}Quadrants II and IV}~
One can see from the plots of Quadrants II and IV of Figure \ref{fig:na} that M87* is allowed as a KTN BH for the low value ($|n_*| \leqslant 1$) as well as the high value ($|n_*| > 1$) of $n_*$, unlike Quadrants I and III. In fact, the value of $|n_*|$ of M87* as a KTN BH or NS can be greater than even $2$ (i..e, $|n_*| > 2$), which is not shown in the plot. To be a KTN NS of much lower value of $n_*$ (i.e., $|n_*| < 1$), the allowed ranges for $a_*$ starts from $|a_*| \sim 4.5$.  We plot Figures~\ref{nai17_4q} and \ref{fig:na} for a limited parameter space. For a general large parameter space, much higher values of $|a_*|$ and/or $|n_*|$ are possible for M87* as a KTN BH or a KTN NS in Quadrants II and IV, if we compare it to the other two quadrants (I and III). 
For instance, the black hatched region of NS, which spans with $4.5 \lesssim |a_*| \lesssim 6.2$ (i.e., $\d a_* \sim 1.7$) for $|n_*| \rightarrow 0$, shifts towards right (left) of the plot in Quadrant IV (Quadrant II) with the higher values of $|a_*|$ for increasing the value of $|n_*|$. 
As an example, the black hatched region of NS extends in the following range of $a_* : 19 < a_* < 21$ for $n_*=-10$. However, the width of black hatched region of NS remains almost constant ($\d a_* \sim 2$) for a fixed value of $n_*$ in the following range: $0 < |n_*| < 10$. Not only this, the value of $\Delta d$ also remains almost constant (i.e., $\Delta d \sim 1\%$) for the above-mentioned black hatched region of NS.

As it is seen from the discussion of Sec. \ref{sec:ktn} that, for the general large parameter space,
the thin YC regions, adjacent to the green dashed lines, are broadened for increasing the value of $|n_*|$, and the black hatched region of NS can be overlapped with this region. Note that the value of $\Delta d$ in the black hatched regions of NS is not necessarily always be within $2\%$ as shown in the limited parameter space of Figure \ref{fig:na}, it can also be equal to or greater than $10\%$ (i.e., $\Delta d \lesseqgtr 10\%$) depending on the values of $a_*$ \& $n_*$ in the general large parameter space, specifically close to the green dashed lines. For example, if M87* is a KTN NS of Quadrant IV with $a_* \sim 5.2$ and $n_* \sim -5$, the shadow size would be $\sim 12$ with $\Delta d \sim 10\%$.

\section{Conclusion and discussion}
\label{sec:dis}

The KTN metric is the mathematical solution of the Einstein field equation, and
the EHT observations \cite{EHT5} have not ruled out the possibility
of M87* containing the gravitomagnetic monopole/NUT charge, yet. In this work, we have investigated the
possibility for the existence of gravitomagnetic monopole in M87*, by studying
how the shadow size and shape depends on the KTN metric parameters and using the first observational image of its shadow as constraints. We have found that the observational constraints on the size and circularity of the M87* shadow do not exclude the possibility that this compact object can be a naked singularity and contain the gravitomagnetic monopole. 
% In fact, as the first image of M87* released by the EHT does not seem to be exactly `circular' \cite{EHT4, EHT6}, one may think M87* as a KTN NS considering the result obtained in Sec. \ref{sec:m87}.
It is, therefore, important to have accurate measurements of both the shadow size and asymmetry, which can be used to put strong constraints on $a_*$ and $n_*$, and break the degeneracies between the different metrics as well as between the BHs and NSs. If M87* is really a KTN NS, the accurate measurements of both the shadow size and asymmetry can help to break the degeneracies between the different quadrants. 
% For instance, the circularity condition excludes the possibility of M87* to be a Taub-NUT BH, i.e., a non-zero $n_*$ with $a_* \rightarrow 0$. This can be seen from Figure~\ref{fig:na}, as the shadow should be exactly circular in that particular case. 

Secondly, our finding reveals that the observational image is not only compatible with the KTN
metric of the limited ranges of $a_*$ and $n_*$ as shown in
Figure~\ref{fig:na}, but it is also compatible with the general large parameter space as discussed in Sec. \ref{sec:m87}.
Furthermore, we have also shown that if M87* in fact contains the gravitomagnetic monopole  
the upper limit of $n_*$ cannot be greater than $1.1$, i.e., $n_* \lesssim 1.1$ in case of the prograde rotation ($a_* > 0$), whereas the lower limit
of $n_*$ cannot be less than $-1.1$, i.e., $ n_* \gtrsim -1.1$ in case of the retrograde rotation ($a_* < 0$). If the deviation from the circularity of the shadow is found to be less than $1\%$ by the future EHT-like observations (such as, ngEHT \cite{ng}), the Kerr and KTN NSs are falsifiable. In spite of that, M87* could still be described as a KTN BH with the upper limit of $n_*$ as $n_* < 0.9$ in case of the prograde rotation ($a_* > 0$), and the lower limit of $n_*$ as $ n_* > -0.9$ in case of the retrograde rotation ($a_* < 0$). Note that our conclusion can be true, if M87* as a AGN (see \cite{lnbl}) contains the NUT charge. 
\\

%%%%%%%%%%%%%%%%%%%%%%%%%%%%%%%%%%%%%%%%%%%%%%%%%%%%%%%%%

{\bf Acknowledgments}

This work is partly supported by the National Key Program for Science and
Technology Research and Development (Grant No.~2016YFA0400703, 2016YFA0400704),
the National Natural Science Foundation of China under grant No.~11873056,
11690024, 11673001, 11750110410, 11721303, and the Strategic Priority Program of the
Chinese Academy of Sciences (Grant No.~XDB 23040100). M.\ G.-N. and C. C. acknowledge
support from the China Postdoctoral Science Foundation, Grant No.~2017LH021 and ~2018M630023 respectively.
%%%%%%%%%%%%%%%%%%%%%%%%%%%%%%%

\begin{appendix}

\section{\label{app}Relation between $i$ and $\th_s$ in the Kerr-Taub-NUT spacetime}

{\bf General case:} Eq. (\ref{sing}) shows that although the singularity is always located at the equatorial plane: $\th_s=\pi/2$ in the Kerr spacetime, this is not true for the KTN spacetime. Therefore, one can write Eq. (\ref{sing}) as
 \begin{eqnarray}
  a_*=-n_*\sec \th_s
  \label{eq1}
 \end{eqnarray}
Now, substituting this (Eq. \ref{eq1}) into the basic NS condition:
  \begin{eqnarray}
   a_*^2 > 1+n_*^2
  \end{eqnarray}
we obtain
 \begin{eqnarray}
 \pm n_{*\th_s} &>& \cot \th_s ,
 \label{eq3}
 \\
   {\rm i.e., } \,\,\,\, n_{*\th_s} &>& \cot \th_s \,\,\,\, (\rm for \,\, Quadrant \,\, II)
  \\
 {\rm and, } \,\,\,\, n_{*\th_s} &<& -\cot \th_s
 \,\,\,\, (\rm for \,\, Quadrant \,\, IV).
 \end{eqnarray}
 The corresponding value of $a_*$ can be obtained from Eq. (\ref{eq1}) as :
 \begin{eqnarray}
a_{*\th_s} &<-& {\rm cosec}~\th_s \,\,\,\, (\rm for \,\, Quadrant \,\, II)
\\
 {\rm and, } \,\,\,\, a_{*\th_s} &>& {\rm cosec}~\th_s
 \,\,\,\, (\rm for \,\, Quadrant \,\, IV).
 \end{eqnarray}
Below we discuss only for Quadrant IV, as the discussion on Quadrant II is similar to this one. For a fixed value of $n_*$ in Figure \ref{nai17_4q}, the shadow size starts to decrease rapidly if one crosses $a_*=-n_*\sec \th_s$ (not for crossing $a_*=\sqrt{1+n_*^2}$). This is valid for those values of $n_*$ which satisfy: $n_* < -\cot\th_s$. In contrast, the shadow size suddenly becomes very small due to crossing $a_*=\sqrt{1+n_*^2}$ (which is similar to the Kerr case) for those values of $n_*$ which satisfy: $0 > n_* > -\cot \th_s$. This term ($-n_*\sec \th_s$) does not show any interesting effect for the later case, as its value falls in the BH region, i.e., $-n_*\sec \th_s \leqslant \sqrt{1+n_*^2}$, but it plays an important role if its value falls in the NS region, i.e., $-n_*\sec \th_s > \sqrt{1+n_*^2}$. We have already calculated that this term comes outside of the BH region from this  particular point ($\zeta_{\th_s}$)\footnote{$\zeta_{\th_s}$ also satisfies the condition of the extremal BH, i.e., the horizon is located at $R_h=M\left(1+\sqrt{1+(-\cot \th_s)^2-({\rm cosec}~ \th_s)^2}\right)=M$}
\begin{eqnarray}
 \zeta_{\th_s}\equiv(a_{*\th_s}, n_{*\th_s})=({\rm cosec}~ \th_s, -\cot \th_s).
 \label{zeta}
\end{eqnarray}
This means that one can see its effect beyond this particular point and in that case this particular term ($-n_*\sec \th_s$) plays an important role for the size of the shadow. In this special case, the shadow of KTN NENSs can be much bigger (comparing to the Kerr NENSs) for the following range of $a_*: \sqrt{1+n_*^2} < a_* \leq -n_*\sec \th_s$ but it starts to decrease rapidly for crossing the point $a_* = -n_*\sec \th_s$ for a fixed $n_*$ and eventually becomes very small. After that the shadow size increases  again.
Due to the same reason, the shadow size does not appear to be smaller for the transition from the extremal BH to NENS for $n_* < -\cot \th_s$. Thus, if one moves from  $a_*=\sqrt{1+n_*^2}$ (extremal BH) towards right along the $OX$ axis with a fixed $n_*$, one cannot see any abnormality in the shadow size for the transition from the extremal BH to NENS. The violet colored region adjacent to the green dashed line of Figure \ref{nai17_4q} is, therefore, shifted towards the right side of the plot. This special transition (shadow starts to decrease rapidly) occurs at $(a_*, n_*) \equiv (-n_* \sec\th_s, n_*)$ for a fixed $n_*$. 

\subsection*{Application to the recent EHT result for M87*}
The above mentioned scenario can be discussed considering an example of the recent EHT result for M87*. In this case, the transition starts at the point (see Eq. \ref{zeta})
\begin{eqnarray}
 \zeta_{17}=(a_{*17}, n_{*17})=({\rm cosec}~ 17^{\circ}, -\cot 17^{\circ}) \equiv (3.421, -3.271) \nonumber
 \\ 
 \end{eqnarray}
 for $\th_s=17^{\circ}$ in Quadrant IV. 
Therefore, the violet colored region of Figure \ref{nai17_4q} will be shifted towards the right side of the plot for $n_* \lesssim -3.271$ (specifically from this point: $\zeta_{17} = (3.421, -3.271)$). For this case, if we move from the green dashed line of Figure \ref{nai17_4q} towards right (i.e., $a_* \rightarrow 7$) along a fixed $n_*$, one cannot see any abnormality in the shadow size for the transition from the extremal BH to NENS. However, the violet colored region still remains adjacent to the green line for $0 > n_* > -3.271$. The special transition (shadow starts to be smaller rapidly) occurs at $(a_*, n_*) \equiv (-n_* \sec 17^{\circ}, n_*)=(-1.046n_*, n_*)$ for a fixed $n_*$. The whole picture is also clear from the values given in Table \ref{tb1}.
 
 \begin{table*}{}
 \begin{center}
 \begin{tabular}{| l | c | r | l | c | l |}
  \hline                        
$n_*$  & $\sqrt{1+n_*^2}$ &$-n_*\sec 17^{\circ}$ &$a_*$ & $\Delta d$ & Shadow size
\\ 
\hline 
   & &  & 1.00 & 0.002 & 10.60 \\
-1 & 1.41 & 1.05 & 1.40 & 0.012 & 8.73   \\
      & &  & 1.50 & 0.050 & 0.567  \\  
\hline 
    & &  & 2.00 & 0.005 & 10.95 \\
-2 & 2.24 & 2.09 & 2.20 & 0.015 & 9.38 \\
     & &  & 2.30 & 0 & 0.703 \\ 
\hline                    
-3.27 &  3.42 & 3.42 & 3.40 & 0.024  &  10.18 \\
      & & &  3.44 ($18.09^{\circ}=\th_s > i$) & 0 & 0.003 \\ 
\hline    
& & &4.12  &  0.038 & 10.38 \\
& & &4.16 ($15.94^{\circ}=\th_s < i$) &  0.096 & 11.23 \\
-4   & 4.12 & 4.18  & 4.18 ($17.00^{\circ}=\th_s = i$)  & 0.025 &   7.43 \\
      & & & 4.20 ($17.75^{\circ}=\th_s > i$) &0   & 0.003\\ 
\hline                      
-6 & 6.08 & 6.27 & 6.25 ($16.26^{\circ}=\th_s < i$) & 0.108 & 13.45 \\
  & & &6.29 ($17.47^{\circ}=\th_s > i$) & 0	        & 0.003 \\
\hline  
-8 & 8.06 &8.36 & 8.34 ($16.42^{\circ}=\th_s < i$) & 0.117 &  15.57 \\
    &  & & 8.38 ($17.32^{\circ}=\th_s > i$) & 0 & 0.003 \\ 
\hline                 
            
\end{tabular}
\caption{The shadow size starts to decrease rapidly for crossing $a_*=-n_*\sec 17^{\circ}$ (or, $a_*=-1.046n_*$), not for crossing $a_*=\sqrt{1+n_*^2}$, which is valid for the values of $n_*: n_* < -3.27$. In contrast, the shadow size suddenly becomes very small due to crossing $a_*=\sqrt{1+n_*^2}$ for the values of $n_*: 0 > n_* > -3.27$ (similar to the case for Kerr spacetime). This term ($-n_*\sec 17^{\circ}$) does not show any interesting effect in the shadow sizes, as its value falls in the BH region (see the first two columns for $n_*=-1$ and $-2$),  but the rapid decreasing in the  shadow sizes is seen in the last three columns (see $n_*=-4, -6$ and $-8$) as its ($-n_*\sec 17^{\circ}$) value falls in the NS region. For the latter, the shadow of KTN NENS can be bigger (comparing to the Kerr NENSs) but it starts to decrease rapidly for crossing the point $a_* = -n_*\sec 17^{\circ}$ for a fixed $n_*$, and eventually becomes very small. After that the shadow size increases  again.}
  \label{table}
  \label{tb1}  
   \end{center}  
\end{table*}
Interestingly, a deeper study of the values given in Table \ref{tb1} reveals that, if $\th_s$ is greater than the inclination angle ($i$), i.e., $\th_s > i$, one can see an extremely smaller shadow, whereas the shadow will be much bigger for $\th_s \leqslant i$. The above statement is true in either way, i.e., depending on a slight change in the inclination angle  from the location of singularity, one can see a huge difference in the shadow size of a same KTN object. This does not arise in case of the Kerr spacetime ($n_*=0$), as
\begin{eqnarray}
 \zeta_{90}=(a_{*90}, 0)=({\rm cosec}~ 90^{\circ}, -\cot 90^{\circ})=(1, 0)
 \end{eqnarray}
 is a constant for any value of $i$. Therefore, the rapid decreasing in the shadow sizes always occurs  at $|a_*|=1$, during the transition from the extremal Kerr BH to the Kerr NENS for any inclination angle $i$. 

Here, we should note that Eqs. (\ref{eq1}) and (\ref{eq3}) hold for both the Quadrants II \& IV and this special situation arises only in these two quadrants, for which the location of singularity  can vary from $\th_s \rightarrow \pi/2$  to $\th_s =0$, as discussed in Sec. \ref{sec:1}.
 
\end{appendix}

%%%%%%%%%%%%%%%%%%%%%%%%%%%%%%%


\begin{thebibliography}{}

\bibitem{EHT1} The Event Horizon Telescope Collaboration,
Astrophys. J. {\bf 875}, L1 (2019).
  
  \bibitem{EHT5} The Event Horizon Telescope Collaboration, 
Astrophys. J. {\bf 875}, L5 (2019).

 \bibitem{nut} E. Newman, L. Tamburino, T. Unti, J. Math. Phys. {\bf 4}, 915 (1963).
 
 \bibitem{ckj} C. Chakraborty, P. Kocherlakota, P. S. Joshi, 
Phys. Rev.
{\bf D 95}, 044006 (2017).

 \bibitem{ckp} C. Chakraborty, P. Kocherlakota, M. Patil, S. Bhattacharyya,
P. S. Joshi, A. Kr\'olak, Phys. Rev. {\bf D 95}, 084024 (2017).

\bibitem{sc} C.~Bambi, K.~Freese, S.~Vagnozzi and L.~Visinelli,
Phys. Rev. {\bf D 100}, 044057 (2019).

\bibitem{jh} J. B. Hartle, {\it Gravity:An introduction to 
Einstein's General relativity}, Pearson (2009).

\bibitem{wei} S-W. Wei, Y-X. Liu, C-E Fu, K. Yang, JCAP {\bf 10} (2012) 053.


\bibitem{rs2}  S. Ramaswamy, A. Sen, Phys. Rev. Lett. {\bf 57}, 1088 (1986). 

\bibitem{rs}  S. Ramaswamy, A. Sen, J. Math. Phys. (N.Y.) {\bf 22}, 2612 (1981). 

\bibitem{bon} W. B. Bonnor, Proc. Camb. Phil. Soc. {\bf 66}, 145 (1969).

\bibitem{dow} J. S. Dowker, Gen. Rel. Grav. {\bf 5}, 603 (1974).


 \bibitem{mis} C. W. Misner, J. Math. Phys. {\bf 4}, 924 (1963).
 
 \bibitem{hehl} F. W. Hehl,  P. v. d. Heyde, G. D. Kerlick, Rev. Mod. Phys. {\bf 48}, 393 (1976). 
 
 \bibitem{bini} D. Bini et al., Class. Quantum Grav. {\bf 20}, 457 (2003).

\bibitem{zs} R. L. Zimmerman, B. Y. Shahir, Gen. Rel. Grav. {\bf 21}, 821 (1989).

\bibitem{pnas} C. Bunster, M. Henneaux, PNAS {\bf 104}, 12243  (2007).

\bibitem{bun} C. Bunster et al., Phys. Rev. {\bf D 73}, 105014 (2006).

\bibitem{vir} A. Virmani, Phys. Rev. {\bf D 84}, 064034 (2011).

% \bibitem{rt} T. Regge, C. Teitelboim, Ann. Phys. (N.Y.) {\bf 88}, 286 (1974).


 \bibitem{kag} V. Kagramanova et. al, Phys. Rev. {\bf D 81}, 124044 (2010).
 
\bibitem{nov} I. D. Novikov, {\it Evolution of the Universe},
Cambridge University Press, New York (1983)


\bibitem{fn} J. Friedman et al., Phys. Rev. {\bf D 42}, 1915 (1990)

\bibitem{lnbl} D. Lynden-Bell, M. Nouri-Zonoz, Rev. Mod. Phys. {\bf 70}, 427 (1998). 

 \bibitem{liu} C. Liu, S. Chen, C. Ding, J. Jing, Phys. Lett. {\bf B 701}, 285 (2011).

% \bibitem{cc2} C. Chakraborty, Eur. Phys. J. {\bf C 74}, 2759 (2014). 

\bibitem{cc} C. Chakraborty, Eur. Phys. J. {\bf C 75}, 572 (2015). 

\bibitem{cbgm} C. Chakraborty, S. Bhattacharyya, Phys. Rev. {\bf D 98}, 043021 (2018).

\bibitem{cbgm2} C. Chakraborty, S. Bhattacharyya, JCAP {\bf 05} (2019) 034.

\bibitem{1Amir} M.~Amir, K.~Jusufi, A.~Banerjee, S.~Hansraj, Class. Quant. Grav. \textbf{36}, 215007 (2019).

\bibitem{3Amir} M.~Amir, A.~Banerjee and S.~D.~Maharaj, Annals Phys. \textbf{400}, 198 (2019).


\bibitem{4Jusufi} K.~Jusufi, M.~Amir, M.~S.~Ali, S.~D.~Maharaj, Phys. Rev. D \textbf{102}, 064020 (2020).

\bibitem{2Ghosh}
S.~G.~Ghosh, M.~Amir, S.~D.~Maharaj, Nucl. Phys. B \textbf{957}, 115088 (2020).

\bibitem{6Kumar} R.~Kumar, S.~G.~Ghosh, A.~Wang, Phys. Rev. D \textbf{100}, 124024 (2019).

\bibitem{5Kumar} R.~Kumar, S.~G.~Ghosh, JCAP \textbf{07}, 053 (2020).

\bibitem{7Kumar} R.~Kumar, S.~G.~Ghosh, Astrophys. J. \textbf{892}, 78 (2020).

\bibitem{afrin} M.~Afrin, R.~Kumar and S.~G.~Ghosh, Mon. Not. Roy. Astron. Soc. \textbf{504}, 5927 (2021).


\bibitem{ml} J. G. Miller, J. Math. Phys. {\bf 14}, 486 (1973).

\bibitem{mr} V. S. Manko, E. Ruiz, Class. Quantum Grav. {\bf 22}, 3555 (2005).

\bibitem{mcd} S. Mukherjee, S. Chakraborty, N. Dadhich, Eur. Phys. J. {\bf C 79}, 161 (2019).

\bibitem{myshadow1} M.~Ghasemi-Nodehi, Z.~Li and C.~Bambi, Eur. Phys. J. {\bf C 75}, 315 (2015).
  
 \bibitem{myshadow2} M.~Ghasemi-Nodehi, C.~Bambi, Eur. Phys. J. {\bf C 76}, 290 (2016).

\bibitem{Lund:2009zzb} E.~Lund, L.~Bugge, I.~Gavrilenko and A.~Strandlie, JINST {\bf 4}, P04001 (2009).
 
 \bibitem{js2} T. Johannsen, D. Psaltis, Astrophys. J. {\bf 718}, 446 (2010).
 
 \bibitem{coba} C.~Bambi, Astrophys. J. {\bf 761}, 174 (2012).
 
  \bibitem{tak} R. Takahashi, Astrophys. J. {\bf 611}, 996 (2004).
 
  \bibitem{js1} T. Johannsen, D. Psaltis, Astrophys. J. {\bf 716}, 187 (2010).
 
 
 \bibitem{chastu} D. Charbulak, Z. Stuchlik, Eur. Phys. J. {\bf C 78}, 879 (2018).

\bibitem{ch} S. Chandrashekar, {\it The Mathematical Theory of Black Holes}, Clarendon Press, Oxford (1983).


 \bibitem{abdu} A. Abdujabbarov et al., 
Astrophys. Space. Sci. {\bf 344}, 429 (2013).
 
 \bibitem{gh} E. G. Gimon, P. Horava, Phys. Lett. {\bf B 672}, 299 (2009).


\bibitem{EHT6} 
 The Event Horizon Telescope Collaboration, Astrophys. J. {\bf 875}, L6 (2019).

\bibitem{jet} R.~Craig Walker, P.~E.~Hardee, F.~B.~Davies, C.~Ly and W.~Junor, Astrophys. J.  {\bf 855}, 128 (2018).


\bibitem{ng} The Next Generation Event Horizon Telescope (ngEHT):  https://pweb.cfa.harvard.edu/news/announcement-next-generation-event-horizon-telescope-design-program



\end{thebibliography}
\end{document}